\documentclass{article}
\usepackage{t1enc}
\usepackage[utf8]{inputenc}
\usepackage[english]{babel}
\usepackage{graphicx}
\usepackage{amssymb}
\usepackage{amsmath}
\usepackage{amsthm}

\def\OA{{\cal A}}
\def\DK{{\cal O}}
\def\SDK{{\cal K}}
\def\PR{{\cal P}}
\def\vNA{{\cal N}}
\def\Mink{{\cal M}}
\def\CO{{\bf C}}
\def\CO{{\bf C}}
\def\RE{{\bf R}}
\def\IN{{\bf Z}}

\def\fel{{\frac{1}{2}}}

\def\UN{\mathbf{1}}
\def\qed{\ \vrule height 5pt width 5pt depth 0pt}
\def\cros{\raise1.9pt\hbox{$\scriptscriptstyle
          >$}\!\raise1.5pt\hbox{$\scriptstyle\triangleleft\,$}}
\theoremstyle{definition}\newtheorem{D}{Definition}
\theoremstyle{definition}
\theoremstyle{definition}\newtheorem{Prop}{Proposition}
\theoremstyle{definition}\newtheorem{Lemma}{Lemma}

\title{\bf Bell inequality and common causal explanation \\ in algebraic quantum field theory}
\author{\textit{G\'abor Hofer-Szab\'o}\thanks{King Sigismund College, Budapest, email: gsz@szig.hu} \\
\textit{P\'eter Vecserny\'es}\thanks{Wigner Research Centre for Physics, Budapest, email: vecsernyes.peter@wigner.mta.hu}}

\date{ }

\begin{document}
\maketitle

\begin{abstract}
Bell inequalities, understood as constraints between classical conditional probabilities, can be derived from a set of assumptions representing a common causal explanation of classical correlations. A similar derivation, however, is not known for Bell inequalities in algebraic quantum field theories establishing constraints for the expectation of specific linear combinations of projections in a quantum state. In the paper we address the question as to whether a `common causal justification' of these non-classical Bell inequalities is possible. We will show that although the classical notion of common causal explanation can readily be generalized for the non-classical case, the Bell inequalities used in quantum theories cannot be derived from these non-classical common causes. Just the opposite is true: for a set of correlations there can be given a non-classical common causal explanation \textit{even if} they violate the Bell inequalities. This shows that the range of common causal explanations in the non-classical case is wider than that restricted by the Bell inequalities.
\vspace{0.1in}

\noindent
\textbf{Key words:} Bell inequality, common cause, noncommutativity, algebraic quantum field theory.
\end{abstract}

\section{Introduction}

The original context which led to the formulation of the Bell inequalities was the intention to accomodate quantum correlations in a \textit{locally causal} theory. The clearest formulation of such a theory is due to Bell himself (Bell, 1987, p. 54). In a number of seminal papers Bell carefully analyzed the intuitions lying behind our notion of locality and causality. His major contribution, however, consisted in translating these intricate notions into a simple probabilistic language which made these notions tractable both for mathematical treatment and later for experimental testability. This probabilistic framework made it possible to exactly identify the probabilistic requirements responsible for the violation of the Bell inequalities in the EPR scenario. A decade later authors like Van Fraassen (1982), Jarrett (1984) and Shimony (1986) spent much time to analyze the philosophical consequences of giving up either the one or the other of these probabilistic assumptions. It also turned out soon that the conceptual framework in which the Bell inequalities can be treated most naturally is the common causal explanation of correlations, originally stemming from Reichenbach (1956) and later adopted to the EPR case by Van Fraassen (1982).

Since the aim of these considerations was to accomodate the EPR scenario in a classical world picture, both Bell and the subsequent writers used a \textit{classical} probabilistic framework in their analysis. All the assumptions representing locality and causality and also the resulting Bell inequalities were formulated in the language of the classical probability theory. Now, if the Bell inequalities were classical, how could they be violated in the EPR scenario which is well known to be described by quantum theory? Well, the answer is that quantum theory with its mathematical structure and ontological commitments played \textit{no role at all} in the Bell scenario. Quantum mechanics was only used \textit{to generate classical probabilities}, more specifically, \textit{classical conditional probabilities} by the Born rule. These classical conditional probabilities, however, could also have been gained directly from the experiments, and indeed later they have been gained so. In other words, the original context of the Bell inequalities has no intimate link to quantum theory even if quantum theory produces probabilities which, reinterpreted as classical conditional probabilities, violate those inequalities. This classical view on the Bell inequalities manifests itself in various authors. Nicolas Gisin for example writes: ``Bell inequalities are relations between conditional probabilities valid under the locality assumption.'' (Gisin 2009, p. 126)

In the face of all these, the Bell inequality has made its way into quantum theory. It has been soon formulated as a general mark of entanglement of the given quantum state on a $C^*$-algebra  (Summers and Werner 1987a, b). A quote from Bengtson and Zyczkowski (2006, p. 362) might illustrate this change of focus in the role of Bell inequalities: ``The Bell inequalities may be viewed as a kind of separability criterion, related to a particular entanglement witness, so evidence of their violation for certain states might be regarded as an experimental detection of quantum entanglement.'' How could the Bell inequality make its way to this non-classical formalism so alien from its original context? Does there exist a justification for this `trespass'?

In this paper we would like to investigate a possible justification for this transition. In this justification we intend to follow the route pioneered by Bell, Van Fraassen, Jarrett, Shimony and others in that we stick to the conviction that the Bell inequalities follow from the requirement of implementing correlations into a \textit{locally causal} theory. We transcend, however, this view in \textit{not} assuming that this theory has to be \textit{classical}. Or in other words, we pose the question whether the probabilistic requirements representing local causality and constituting the core of the Bell inequalities can be reasonable formulated also in a non-classical theory. 

A natural candidate for such a non-classical theory with clear conceptions of locality and causality is algebraic quantum field theory (AQFT) (Haag, 1992). In AQFT events are represented by projections with well defined spacetime support and local causality is ensured by a set of axioms. Hence we can pose the question as to whether the Bell inequalities featuring in AQFT follow from a locally causal explanation of correlations in a similar manner to the classical case. Since we intend to give a causal explanation for \textit{correlations between events}, therefore causal explanation is meant to be a \textit{common causal} explanation. We will see that the connection between a common causal explanation and the Bell inequalities in AQFT is not so tight as in the classical case. In the classical case common causes necessarly commute (in the set theoretical `meet' operation) with their effects. Since the quantum events of AQFT form a noncommutative structure, one can decide whether to require that common causes commute with their effects or not. If commutativity is required, the Bell inequalities will follow from the common cause just like in the classical case. But, as we will argue, requiring commutativity is only a remininscence of the classical treatment of correlations and is completely unjustified in the quantum case (see e.g. (Clifton, Ruetsche 1999)). For noncommuting common causes the Bell inequalities will turn out \textit{not} to be derivable from the presence of the common cause---at least not in the similar way to the the classical derivation. This raises the question whether correlations violating the Bell inequalities can have a noncommuting common causal explanation. We will answer this question in the affirmative showing up a situation when a set of correlations maximally violating a specific type Bell inequality has a common causal explanation, which is local in the sense that it can be accomodated in the intersection of the causal pasts of the correlating events. The model we use for this example is the local quantum Ising model, the simplest AQFT with locally finite degrees of freedom.

The paper is structured as follows. In Section 2 we briefly collect the most important concepts and some of the representative propositions concerning the Bell inequality in AQFT. In Section 3 and 4 we give the definition of the classical and the non-classical common causal explanations, respectively, and show how these explanations relate to the Bell inequalities. Since the correct `translation' of the so-called locality and no-conspiracy conditions of the classical common causal explanation into the non-classical setting is a subtle point not needed for our main purpose, we transfer it into the Appendix. Now, the common causal explanations in the EPR-Bell scenario is always meant as providing a \textit{joint} common cause for a \textit{set} of correlations. Providing a joint common cause for a set of correlations is much more demanding than simply providing a common cause for a \textit{single} correlation. Therefore in Section 5, preparing for the more complicated case, we investigate the possibility of a common causal explanation of a single correlation, or in the philosophers' jargon, the status of the Common Causal Principle in AQFT. In Section 6 we return to our original question and present a noncommutative common causal explanation for a \textit{set} of correlations maximally violating some Bell inequalities. In Section 7 we briefly analyze the philosophical consequences of applying \textit{noncommuting} common causes in our causal explanation. We conclude the paper in Section 8.

\section{The Bell inequality in algebraic quantum field theory}

In this Section we collect the most important concepts and some of the representative propositions concerning the Bell inequality in AQFT (see (Summers 1990) and (Halvorson 2007)). We start with the general $C^*$-algebraic setting and then go over to the special algebraic quantum field theoretical formulation.

In the general $C^*$-algebraic setting Bell inequality is treated in the following way. Let $\mathcal A$ and $\mathcal B$ be two mutually commuting $C^*$-subalgebras of some $C^*$-algebra $\mathcal C$. A \textit{Bell operator} $R$ for the pair ($\mathcal A, \mathcal B$) is an element of the following set:
\begin{eqnarray*}
\mathbb{B}(\mathcal A, \mathcal B) & := & \left\lbrace \frac{1}{2} \big( X_1(Y_1 + Y_2) + X_2(Y_1 - Y_2) \big) \, {\big |} \, X_i =X_i^* \in \mathcal A; \, Y_i =Y_i^* \in \mathcal B; \, -\UN \leqslant X_i, Y_i  \leqslant \UN \right\rbrace 
\end{eqnarray*}
where $\UN$ is the unit element of $\mathcal C$. For any Bell operator $R$ the following can be proven: 
\begin{description}
\item[Theorem 1.] For any state $\phi\colon\mathcal C\to\CO$, one has $| \phi(R) | \leqslant \sqrt{2}$.
\item[Theorem 2.] For separable states (i.e. for convex combinations of product states) $| \phi(R) | \leqslant 1$.
\end{description}
The \textit{Bell correlation coefficient} of a state $\phi$ is defined as 
\begin{eqnarray*}
\beta(\phi, \mathcal A, \mathcal B) & := & \sup \left\lbrace | \phi(R) | \, {\big |} \, R \in \mathbb{B}(\mathcal A, \mathcal B)  \right\rbrace 
\end{eqnarray*}
and the \textit{Bell inequality} is said to be \textit{violated} if $\beta(\phi, \mathcal A, \mathcal B) > 1$, and \textit{maximally violated} if $\beta(\phi, \mathcal A, \mathcal B)= \sqrt{2}$. An important result of Bacciagaluppi (1994) is the following:
\begin{description}
\item[Theorem 3.] If $\mathcal A$ and $\mathcal B$ are $C^*$-algebras, then there are some states violating the Bell inequality for $\mathcal A \otimes \mathcal B$ iff both $\mathcal A$ and $\mathcal B$ are non-abelian.
\end{description}
Going over to von Neumann algebras Landau (1987) has shown that the maximal violation of the Bell inequality is generic in the following sense:
\begin{description}
\item[Theorem 4.] Let $\mathcal N_1$ and $\mathcal N_2$ be von Neumann algebras, and suppose that $\mathcal N_1$ is abelian and $\mathcal N_1 \subseteq \mathcal N'_2$ ($\mathcal N'$ being the commutant of $\mathcal N$). Then for any state $\beta(\phi, \mathcal A, \mathcal B) \leqslant 1$. On the other hand, if both $\mathcal N_1$ and $\mathcal N_2$ are non-abelian von Neumann algebras such that $\mathcal N_1 \subseteq \mathcal N'_2$, and if ($\mathcal N_1, \mathcal N_2$) satisfies the 
\textit{Schlieder-property},\footnote{The commuting pair $(\mathcal A,\mathcal B)$ of $C^*$-subalgebras in $\mathcal C$ obeys the Schlieder-property, if for $0\not= A\in\mathcal A$ and $0\not= B\in\mathcal B$, $AB\not=0$. Since in case of von Neumann algebras $A$ and $B$ can be required to be projections, Schlieder-property is the analogue of logical independence in classical logic.} then there is a state $\phi$ for which $\beta(\phi, \mathcal A, \mathcal B) = \sqrt{2}$.
\end{description}
Adding further constraints on the von Neumann algebras one obtains other important results such as the following two:
\begin{description}
\item[Theorem 5.] If $\mathcal N_1$ and $\mathcal N_2$ are \textit{properly infinite}\footnote{The center contains no finite projections.} von Neumann algebras on the Hilbert space $\mathcal H$ such that $\mathcal N_1 \subseteq \mathcal N'_2$, and ($\mathcal N_1, \mathcal N_2$) satisfies the Schlieder-property, then there is a dense set of vectors in $\mathcal H$ inducing states which violate the Bell inequality across $(\mathcal N_1,\mathcal N_2)$ (Halvorson and Clifton, 2000).
\item[Theorem 6.] Let $\mathcal H$ be a separable Hilbert space and let $\mathcal R$ be a von Neumann factor of type $III_1$ acting on $\mathcal H$. Then every normal state $\phi$ of $\mathcal B(\mathcal H)$ maximally violates the Bell inequality across $(\mathcal R,\mathcal R')$ (Summers and Werner, 1988).
\end{description}
Type $III$ factors featuring in Theorems 5-6. are the typical local von Neumann algebras in AQFT with locally infinite degrees of freedom. Here we briefly survey the basic notions of the theory.

In AQFT observables (including quantum events) are represented by unital $C^*$-algebras associated to bounded regions of a given spacetime. The association of algebras and spacetime regions is established along the following lines. 
\begin{itemize}
\item[(i)] \textit{Isotony.} Let $\mathcal{S}$ be a spacetime. A \textit{double cone} in $\mathcal{S}$ is the intersection of the causal past of a point $x$ with the causal future of a point $y$ timelike to $x$. Let $\SDK$ be a collection of double cones of $\mathcal{S}$ such that $(\SDK,\subseteq)$ is a directed poset under inclusion $\subseteq$. The net of local observables is given by the isotone map $\SDK\ni V\mapsto\OA(V)$ to unital $C^*$-algebras, that is $V_1 \subseteq V_2$ implies that $\OA(V_1)$ is a unital $C^*$-subalgebra of $\OA(V_2)$. The \textit{quasilocal observable algebra} $\OA$ is defined to be the inductive limit $C^*$-algebra of the net $\{\OA(V),V\in\SDK\}$ of local $C^*$-algebras. 
\item[(ii)] \textit{Microcausality.} The net $\{\OA(V),V\in\SDK\}$ satisfies microcausality (aka Einstein causality): $\OA(V')'\cap\OA \supseteq \OA(V),V\in\SDK$, where primes denote spacelike complement and algebra commutant, respectively. $\OA(V')$ is the smallest $C^*$-algebra in $\OA$ containing the local algebras $\OA(\tilde V),\SDK\ni\tilde V\subset V'$. 
\item[(iii)] \textit{Covariance.} Let $\mathcal{P}_\SDK$ be the subgroup of the group $\mathcal{P}$ of geometric symmetries of $\mathcal{S}$ leaving the collection $\SDK$ invariant. A group homomorphism $\alpha\colon\mathcal{P}_\SDK\to\textrm{Aut}\,\OA$ is given such that the automorphisms $\alpha_g,g\in\mathcal{P}_\SDK$ of $\OA$ act covariantly on the observable net: $\alpha_g(\OA(V))=\OA(g\cdot V), V\in\SDK$.
\end{itemize}

To the net $\{\OA(V),V\in\SDK\}$ satisfying the above requirements we will refer to as a $\mathcal{P}_\SDK$-\textit{covariant local quantum theory}. If $\mathcal{S}= \Mink$ is the Minkowski spacetime and $\SDK$ is the net of all double cones then $\mathcal{P}_\SDK$ is the Poincar\'e group, and we obtain Poincar\'e covariant algebraic quantum field theories with locally infinite degrees of freedom. Restricting the collection $\SDK$ one can obtain $\mathcal{P}_\SDK$-covariant local quantum theories with locally finite degrees of freedom, for instance our example, the local quantum Ising model (see below). 

A \textit{state} $\phi$ in a local quantum theory is defined as a normalized positive linear functional on the quasilocal observable algebra $\OA$. The corresponding GNS representation $\pi_{\phi}\colon\OA\to\mathcal{B}(\mathcal{H}_\phi)$ converts the net of $C^*$-algebras into a net of $C^*$-subalgebras of $\mathcal{B}(\mathcal{H}_\phi)$. Closing these subalgebras in the weak topology one arrives at a net of local von Neumann observable algebras: $\vNA(V):=\pi_{\phi}(\OA(V))'', V\in\SDK$. 

Von Neumann algebras are generated by their projections, which are called \textit{quantum events} since they can be interpreted as 0-1--valued observables. The expectation value of a projection is the probability of the event that the observable takes on the value 1 in the appropriate quantum state. Two commuting quantum events $A$ and $B$ are said to be \textit{correlating} in a state $\phi$ if 
\begin{eqnarray*}
\phi(AB) &\neq & \phi(A)\phi(B).
\end{eqnarray*}
If the events are supported in spatially separated spacetime regions $V_A$ and $V_B$, respectively, then the correlation between them is said to be \textit{superluminal}. To see that superluminal correlations violating Bell inequalities abound in Poincar\'e covariant algebraic quantum field theories, one has to introduce further requirements on the representations of $\OA$ (see Haag 1992):
\begin{itemize}
\item[(iv)] \textit{Unitary implementability.} There is a strongly continuous unitary representation of the Poincar\'e group, $U\colon\mathcal P\to\mathcal{B}(\mathcal{H}_\phi)$, such that
\begin{eqnarray*}
\pi_\phi(\alpha_g (A)) = U(g)\pi_\phi(A)U(g)^*,\qquad A\in\OA,\ g\in\mathcal P. 
\end{eqnarray*}
\item[(v)] \textit{Vacuum condition.} There is a (up to a scalar) unique vector $\Omega$ in the Hilbert space $\mathcal{H}_0$ corresponding to the vacuum state $\phi_0$ such that $U(g)\Omega = \Omega$ for all $g \in \mathcal P$.
\item[(vi)] \textit{Spectrum condition.} The spectrum of the self-adjoint generators of the strongly continuous unitary representation of the translation subgroup $\RE^4$ of $\mathcal P$ lies in the closed forward light cone.
\item[(vii)] \textit{Weak additivity.} For any nonempty open region $V$, the set of operators $\cup_{g\in \mathbb R^4}\vNA(g\cdot V)$ is dense in $\mathcal{B}(\mathcal{H}_0)$ (in the weak operator topology).
\end{itemize}
Now, under conditions (i)-(vii) the local von Neumann algebras supported in spacelike separated double cones satisfy the Schlieder property (Schlieder, 1969). Therefore Theorem 4 applies to these algebras stating that there is a state maximally violating the Bell inequality across these local algebras. Moreover, if the net is \textit{non-trivial}\footnote{For each double cone $V$, $\OA(V)\neq \mathbb{C}\UN$.}, then the local von Neumann algebras are properly infinite. This makes Theorem 5 applicable to local von Neumann algebras supported in spacelike separated double cones stating that there is a dense set of vectors in $\mathcal H$ inducing states which violate the Bell inequality.

Being properly infinite the von Neumann algebras cannot be of type $I_n$ and $II_1$ but they still can be of type $I_{\infty}$ or $II_{\infty}$ . However, a set of independent results indicates that the local von Neumann algebras are of type $III$, more specifically \textit{hyperfinite}\footnote{The weak closure of an ascending sequence of finite dimensional algebras.} factors of type $III_1$. Buchholz et al. (1987) proved that the local algebras for relativistic free fields are type $III_1$ and it was also shown that one can construct the local von Neumann algebras as a unique type $III_1$ hyperfinite factor from the underlying Wightman theory by adding the assumption of \textit{scaling limit} (see (Fredenhagen (1985)). 

Instead of deriving the type of the von Neumann algebras from more general physical requirements, one also can explicitely add this condition as a new axiom of AQFT: 
\begin{itemize}
\item[(viii)] \textit{The type of the algebras.} For every double cone $V$ the von Neumann algebra $\vNA(V )$ is of type $III_1$.
\end{itemize}
Under conditions (i)-(viii) the local von Neumann algebras supported in spacelike separeted double cones satisfy the assumptions of Theorem 6, therefore every normal state will maximally violate the Bell inequality across pairs of algebras supported in spacelike separated double cones.

Finally, we mention a physically important consequence of Theorem 6:
\begin{description}
\item[Theorem 7.] The vacuum state maximally violates the Bell inequality across the \textit{wedge}\footnote{Poincar\'e transforms of the region $W_R:=\{x \in \mathcal{M} \vert x_1 > \vert x_0\vert\}$.} algebras $(\mathcal N(W), \mathcal N(W)')$. (Summers, Werner 1988).
\end{description}
\vspace{0.1in}

\noindent
As said above, the Bell inequality typically used in AQFT is of the following form:
\begin{eqnarray}\label{qCHSH} 
\left\lvert \phi \big(X_1(Y_1 + Y_2) + X_1(Y_1 - Y_2) \big) \right\rvert \leqslant 2,
\end{eqnarray}
where $X_m\in\vNA(V_A)$ and $Y_n\in\vNA(V_B)$ are self-adjoint \textit{contractions} (that is $-\UN \leqslant X_m, Y_n \leqslant \UN$ for $m,n=1,2$) supported in spatially separated spacetime regions $V_A$ and $V_B$, respectively. This type of Bell inequality is usually referred to as the \textit{Clauser--Horne--Shimony--Holte (CHSH) inequality} (Clauser, Horne, Shimony and Holt, 1969). Sometimes in the EPR-Bell literature another Bell-type inequality is used instead of (\ref{qCHSH}): the \textit{Clauser--Horne (CH) inequality} (Clauser and Horne, 1974) defined in the following way:
\begin{eqnarray}\label{qCH} 
-1 \leqslant \phi (A_1 B_1 + A_1 B_2 + A_2 B_1 - A_2 B_2 - A_1 - B_1 ) 
\leqslant 0,
\end{eqnarray}
where $A_m$ and $B_n$ are \textit{projections} located in $\vNA(V_A)$ and $\vNA(V_B)$, respectively. It is easy to see, however, that the two inequalities are equivalent: in a given state $\phi$ the set $\{(A_m, B_n); m,n=1,2\}$ violates the CH inequality (\ref{qCH}) \textit{if and only if} the set $\{(X_m, Y_n); m,n=1,2\}$ of self-adjoint contractions given by
\begin{eqnarray}
X_m &:=& 2A_m - \UN \label{U_A} \\
Y_n &:=& 2B_n - \UN \label{U_B}
\end{eqnarray}
violates the CHSH inequality (\ref{qCHSH}). Therefore, from now on we will concentrate only on the CH-type Bell inequalities.

In the next two sections we turn to the common causal explanation behind the Bell inequalities. In the next Section we introduce the basic notions of the classical common causal explanation leading to the Bell inequalities; in the subsequent Section we generalize these notions for the quantum case.

\section{Classical common causal explanation}

Let us begin with Hans Reichenbach's (1956) original definition which is historically the first probabilistic characterization of the notion of the common cause. Let $(\Omega, \Sigma ,p)$ be a classical probability measure space and let $A$ and $B$ be two positively correlating events in $\Sigma$:
\begin{eqnarray} \label{corr+}
p(A\wedge B) &>& p(A)\, p(B).
\end{eqnarray}
\begin{D} \label{cc}
An event $C\in\Sigma$ is said to be the \textit{Reichenbachian common cause} of the correlation between events $A$ and $B$ if the following  
conditions hold: 
\begin{eqnarray}
p(A\wedge B|C)&=&p(A|C)p(B|C) \label{cc1}\\
p(A\wedge B|C^{\perp})&=&p(A|C^{\perp})p(B|C^{\perp}) \label{cc2}\\
p(A|C)&>&p(A|C^{\perp})  \label{cc3}\\
p(B|C)&>&p(B|C^{\perp}) \label{cc4}
\end{eqnarray}
where $C^{\perp}$ denotes the orthocomplement of $C$ and $p( \, \cdot \,|\, \cdot \,)$ is the conditional probability defined by the Bayes rule. One refers to equations (\ref{cc1})-(\ref{cc2}) as the \textit{screening-off conditions} and to inequalities (\ref{cc3})-(\ref{cc4}) as the \textit{positive statistical relevancy conditions}.
\end{D}

Reichenbach's definition, however, cannot be applied directly to AQFT for four reasons. First, the positive statistical relevancy conditions restrict one to common causes which increase the probability of their effects; or in other words, they exclude negative causes.  Second, the definition also excludes situations in which the correlation is \textit{not} due to a \textit{single} cause but to a \textit{system} of cooperating common causes. Third, it is silent about the spatiotemporal localization of the events. Fourth and most importantly, it is classical.

\begin{figure}[htbp]
\centerline{\resizebox{14cm}{!}{\includegraphics{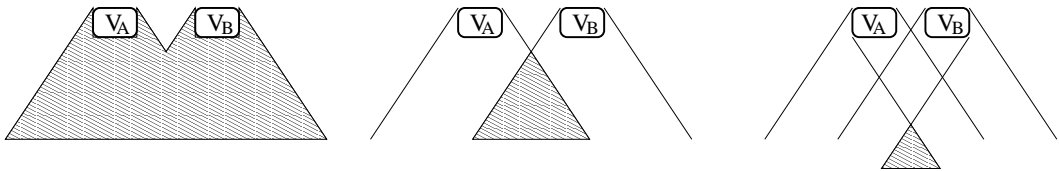}}}
\caption{Possible localizations of the common cause system in different pasts of $V_A$ and $V_B$.}
\label{pastsV.eps}
\end{figure}

Let us first address the first two problems. Let $A$ and $B$ be two correlating events in a classical probability measure space $(\Omega, \Sigma,p)$ that is 
\begin{eqnarray} \label{classicalcorr}
p(A\wedge B) &\neq &p(A)\, p(B).
\end{eqnarray}
\begin{D}\label{ccs}
A partition $\left\{ C_k \right\}_{k\in K}$ in $\Sigma$ is said to be the {\it common cause system} of the correlation (\ref{classicalcorr})
if the following screening-off condition holds for all $k\in K$: 
\begin{eqnarray} \label{ccs1}
p(A \wedge B\vert C_k)&=&p(A\vert C_k)\, p(B \vert C_k),
\end{eqnarray}
where $|K|$, the cardinality of $K$ is said to be the \textit{size} of the common cause system. A common cause system of size $2$ is called a common cause (without the adjective `Reichenbachian', indicating that the inequalities (\ref{cc3})-(\ref{cc4}) are not required).
\end{D}
Concerning the third problem, namely, the localization of the common cause, one has (at least) three different options. Suppose that the two events $A$ and $B$ are localized in two bounded and spatially separated regions $V_A$ and $V_B$ of a spacetime $\mathcal{S}$. Then one can localize $\{C_k\}$ either (i) in the \textit{union} or (ii) in the \textit{intersection} of the causal past of the regions $V_A$ and $V_B$; or (iii) more restrictively, in the spacetime region which lies in the intersection of causal pasts of \textit{every} point of $V_A \cup V_B$. Formally, we have 
\begin{eqnarray*}\label{wcspast}
wpast(V_A, V_B) &:=& I_-(V_A)\cup I_-(V_B) \\
cpast(V_A, V_B) &:=& I_-(V_A)\cap I_-(V_B) \\
spast(V_A, V_B) &:=& \cap_{x \in V_A \cup V_B}\, I_-(x)
\end{eqnarray*}
where $I_-(V)$ denotes the union of the backward light cones i.e. the causal pasts $I_-(x)$ of every point $x$ in $V$ (R\'edei, Summers 2007). We will refer to the above three pasts in turn as the \textit{weak past}, \textit{common past}, and \textit{strong past} of $A$ and $B$, respectively (see Fig. \ref{pastsV.eps}). The notion of these pasts presupposes a spacetime localization structure of the classical event algebra. (For such an attempt see (Henson, 2005).)
\medskip
\medskip

Now, suppose that we do not face \textit{one} correlation $(A,B)$ but a \textit{set} of correlations that is events $A_m$ and $B_n$  in $\Sigma$ such that for any $m \in M,n \in N$
\begin{eqnarray} \label{corrs}
p(A_m\wedge B_n) &\neq& p(A_m)\, p(B_n).
\end{eqnarray}
If our aim is to explain all of these pair-correlations $\{(A_m, B_n); m \in M,n \in N\}$ by a \textit{single} common cause system, then we are led to the following definition:
\begin{D}\label{cccs}
A partition $\left\{ C_k \right\}_{k\in K}$ in $\Sigma$ is said to be a {\it joint\footnote{In (Hofer-Szab\'o and Vecserny\'es, 2012a,b) called \textit{common} common cause system.} common cause system} of the set of correlations $\{(A_m, B_n); m \in M,n \in N\}$ if the following screening-off condition holds for all $m\in M$, $n \in N$, and $k\in K$: 
\begin{eqnarray} \label{cccs1}
p(A_m \wedge B_n\vert C_k)&=&p(A_m\vert C_k)\, p(B_n \vert C_k).
\end{eqnarray}
\end{D}
Obviously, for a set of correlations to have a joint common cause system is much more demanding than to simply have a \textit{separate} common cause system for each correlation.
\vspace{0.1in}

\noindent
Now, let us complicate the picture a little further by introducing \textit{conditional} probabilities. Suppose that events $A_m$ and $B_n$ are \textit{outcomes} of measurements of the observables $\mathrm{A_m}$ and $\mathrm{B_n}$, respectively. Let $a_m$ and $b_n$, respectively denote the events that the appropriate measurement devices are set to measure the observables $\mathrm{A_m}$ and $\mathrm{B_n}$, respectively. Let us refer to these events as \textit{measurement choices}. To be more specific, suppose that each measurement choice $a_m$ in region $V_A$ can yield only two outcomes $A_m$ and $A^{\perp}_m$, and similarly the measurement choices $b_n$ in region $V_B$ can again yield only two outcomes $B_n$ and $B^{\perp}_n$. Finally, suppose that probability of the different measurement choices $a_m$ in region $V_A$ add up to 1, and similarly for the measurement choices $b_n$ in region $V_B$.

Now, the events $A_m$ and $B_n$ are said to be correlating in the \textit{conditional} sense if for all $A_m$, $B_n$, $a_m$, $b_n \in \Sigma$ ($m\in M, n\in N$) the following holds:
\begin{eqnarray} \label{corrcond}
p(A_m\wedge B_n\, \vert \, a_m\wedge b_n) &\neq & p(A_m\vert a_m\wedge b_n)\, p(B_n \vert a_m\wedge b_n) .
\end{eqnarray}
What does a joint common causal explanation of these conditional correlations consists in? The answer to this question is given in the following definition:
\begin{D}\label{lnccccs}
A \textit{local, non-conspiratorial joint common causal explanation} of the conditional correlations (\ref{corrcond}) consists in providing a partition $\{C_k\}$ in $\Sigma$ such that for any $m,m' \in M, n,n' \in N$ the following requirements hold:
\begin{eqnarray}
p(A_m \wedge B_n\vert a_m \wedge b_n \wedge C_k)=p(A_m\vert a_m \wedge b_n \wedge C_k)\,p(B_n \vert a_m \wedge b_n \wedge C_k)  && \mbox{(screening-off)} \label{lncccs1}\\
p(A_m \vert a_m \wedge b_n \wedge C_k)=p(A_m\vert a_m \wedge b_{n'} \wedge C_k) && \mbox{(locality)} \label{lncccs2a} \\ 
p(B_n \vert a_m \wedge b_n \wedge C_k)=p(B_n\vert a_{m'} \wedge b_n \wedge C_k)  && \mbox{(locality)}\label{lncccs2b} \\
p(a_m\wedge b_n \wedge C_k)=p(a_m \wedge b_n)\,p(C_k) && \mbox{(no-conspiracy)}\label{lncccs3}
\end{eqnarray}
\end{D}
The motivation behind requirements (\ref{lncccs1})-(\ref{lncccs3}) is the following. \textit{Screening-off} (\ref{lncccs1}) is simply the application of the notion of common cause for conditional correlations: although $A_m$ and $B_n$ are correlating conditioned on $a_m$ and $b_n$, they will cease to do so if we further condition on  $\{C_k\}$. \textit{Locality} (\ref{lncccs2a})-(\ref{lncccs2b}) is the natural requirement that the measurement outcome on the one side should depend only on the measurement choice on the same side and the value of the common cause but not on the measurement choice on the opposite side. Finally, no-conspiracy (\ref{lncccs3}) is the requirement that the common cause system and the measurement choices should be probabilistically independent. (For the justification of the above requirements by Causal Markov Condition see (Glymour, 2006).)
\vspace{0.1in}

\noindent
Let us now proceed further. A straightforward consequence of Definition \ref{lnccccs} is the following proposition (Clauser, Horne, 1974):
\begin{Prop}\label{PCH} 
Let $A_m$,  $B_n$, $a_m$ and $b_n$ $(m,n=1,2)$ be eight events in a classical probability measure space $(\Omega, \Sigma, p)$ such that the pairs $\{(A_m, B_n); m,n=1,2\}$ correlate in the conditional sense of (\ref{corrcond}). Suppose that $\{(A_m, B_n); m,n=1,2\}$ has a local, non-conspriratorial joint common causal explanation in the sense of Definition \ref{lnccccs}.  Then for any $m,m', n, n' = 1,2; m\neq m'; n\neq n'$ the following \textit{classical Clauser--Horne} inequality holds:
\begin{eqnarray}\label{CH} 
-1 \leqslant p(A_m \wedge B_n \vert a_m \wedge b_n) + p(A_m \wedge B_{n'} \vert a_m \wedge b_{n'}) + p(A_{m'} \wedge B_n \vert a_{m'} \wedge b_n) \nonumber \\
- p(A_{m'} \wedge B_{n'} \vert a_{m'} \wedge b_{n'}) - p(A_m \vert a_m \wedge b_n) - p(B_n \vert a_m \wedge b_n) \leqslant 0
\end{eqnarray}
\end{Prop}

\noindent\textit{Proof.} It is an elementary fact of arithmetic that for any $\alpha , \alpha ', \beta , \beta ' \in [0,1]$ the number
\begin{eqnarray} \label{arith}
\alpha \beta +\alpha \beta '+\alpha '\beta -\alpha '\beta '-\alpha -\beta
\end{eqnarray}
lies in the interval $[-1,0]$. Now let $\alpha , \alpha ', \beta , \beta '$ be the following conditional probabilities:
\begin{eqnarray}
\alpha  &:=& p(A_m \vert a_m \wedge b_n \wedge C_k) \label{a} \\
\alpha ' &:=& p(A_{m'} \vert a_{m'} \wedge b_{n'} \wedge C_k) \label{a'} \\
\beta  &:=& p(B_n \vert a_m \wedge b_n \wedge C_k) \label{b} \\
\beta ' &:=& p(B_{n'} \vert a_{m'} \wedge b_{n'} \wedge C_k)\label{b'}
\end{eqnarray}
Plugging (\ref{a})-(\ref{b'}) into (\ref{arith}) and using locality (\ref{lncccs2a})-(\ref{lncccs2b}) one obtains
\begin{eqnarray}
-1 \leqslant p(A_m \vert a_m \wedge b_n \wedge C_k)p(B_n \vert a_m \wedge b_n \wedge C_k) + p(A_m \vert a_m \wedge b_{n'} \wedge C_k) p(B_{n'} \vert a_m \wedge b_{n'} \wedge C_k) \nonumber \\
+ p(A_{m'} \vert a_{m'} \wedge b_n \wedge C_k)p(B_n \vert a_{m'} \wedge b_n \wedge C_k) - p(A_{m'} \vert a_{m'} \wedge b_{n'} \wedge C_k)p(B_{n'} \vert a_{m'} \wedge b_{n'} \wedge C_k) \nonumber \\
- p(A_m \vert a_m \wedge b_n \wedge C_k) - p(B_n \vert a_m \wedge b_n \wedge C_k) \leqslant 0
\end{eqnarray}
Using screening-off (\ref{lncccs1}) one gets
\begin{eqnarray}\label{hiv1} 
-1 \leqslant p(A_m \wedge B_n \vert a_m \wedge b_n \wedge C_k) + p(A_m \wedge B_{n'} \vert a_m \wedge b_{n'} \wedge C_k) + p(A_{m'} \wedge B_n \vert a_{m'} \wedge b_n \wedge C_k) \nonumber \\
- p(A_{m'} \wedge B_{n'} \vert a_{m'} \wedge b_{n'} \wedge C_k) - p(A_m \vert a_m \wedge b_n \wedge C_k) - p(B_n \vert a_m \wedge b_n \wedge C_k) \leqslant 0
\end{eqnarray}
Multiplying the above inequality by $p(C_k)$, using no-conspiracy (\ref{lncccs3}) and summing up for the index $k$ one obtains
\begin{eqnarray}\label{hiv2} 
-1 \leqslant \sum_k {\big (}  p(A_m \wedge B_n \wedge C_k \vert a_m \wedge b_n) + p(A_m \wedge B_{n'} \wedge C_k \vert a_m \wedge b_{n'}) + p(A_{m'} \wedge B_n \wedge C_k \vert a_{m'} \wedge b_n) \nonumber \\
- p(A_{m'} \wedge B_{n'} \wedge C_k \vert a_{m'} \wedge b_{n'}) - p(A_m \wedge C_k \vert a_m \wedge b_n) - p(B_n\wedge C_k\vert a_m\wedge b_n ) {\big )} \leqslant 0
\end{eqnarray} 
Finally, applying the theorem of total probability
\begin{eqnarray*}
\sum_k p(Y \wedge C_k) = p(Y)
\end{eqnarray*}
one arrives at  (\ref{CH}) which completes the proof.\qed

Proposition \ref{PCH} plays a crucial role in understanding the CH inequality (\ref{CH}). It provides, so to say, a `classical common causal justification' of the classical CH inequality by showing that (\ref{CH}) is a necessary condition for the existence of a local, non-conspriratorial joint common causal explanation for a set of conditional correlations.
\vspace{0.1in}

\begin{figure}[htbp]
\centering{}\includegraphics[width=0.55\columnwidth]{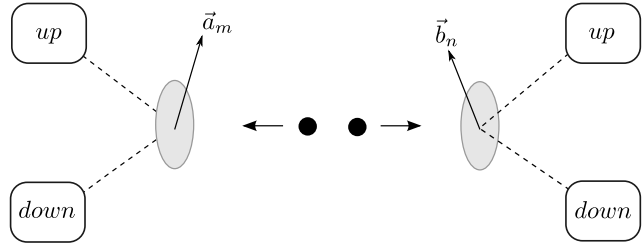}
\caption{EPR--Bohm setup for spin-$\frac{1}{2}$ particles}
\label{epr-bohm-mn}
\end{figure}

\noindent
The well-known situation in which the classical CH inequality (\ref{CH}) is violated and hence the correlations in question have no local, non-conspriratorial joint common causal explanation, is the EPR-Bohm scenario. Consider a pair of spin-$\frac{1}{2}$ particles prepared in the singlet state (see Fig. \ref{epr-bohm-mn}). 
Let $a_{m}$ ($m=1,2$) denote the event that the measurement apparatus is set to measure the spin in direction $\vec{a}_{m}$ in the left wing; and let $p(a_{m})$ stand for the probability of $a_{m}$. Let $b_{n}$ ($n=1,2$) and $p(b_{n})$ respectively denote the same for direction $\vec{b}_{n}$ in the right wing. (Note that $m=n$ does not mean that $\vec{a}_{m}$ and $\vec{b}_{n}$ are parallel directions.) Furthermore, let $p(A_{m})$ stand for the probability that the spin measurement in direction $\vec{a}_{m}$ in the left wing yields the result `up' and let $p(B_{n})$ be defined in a similar way in the right wing for direction $\vec{b}_{n}$. According to the statistical algorithm of quantum mechanics the conditional probability of getting an `up' result \textit{provided} we measure the spin in direction $\vec{a}_{m}$ in the left wing; getting an `up' result \textit{provided} we measure the spin in direction $\vec{b}_{n}$ in the right wing; and getting `up-up' result \textit{provided} we measure the spin in both directions $\vec{a}_{m}$ and $\vec{b}_{n}$ are given by the following relations:
\begin{eqnarray}
p(A_m \vert a_m\wedge b_n) &=& \frac{1}{2} \label{class1} \\
p(B_n\vert a_m\wedge b_n) &=& \frac{1}{2} \label{class2} \\
p(A_m \wedge B_n\vert a_m\wedge b_n) &=& \frac{1}{2} \sin^2\left(\frac{\theta _{a_m b_n}}{2}\right) \label{class3}
\end{eqnarray}
where $\theta _{a_m b_n}$ denotes the angle between directions $\vec{a}_{m}$ and $\vec{b}_{n}$. For non-perpendicular directions $\vec{a}_{m}$ and $\vec{b}_{n}$ (\ref{class1})-(\ref{class3}) predict conditional correlations specified in (\ref{corrcond}). Now, in order to provide a \textit{classical} local, non-conspiratorial joint common causal explanation for these correlations, the conditional probabilities (\ref{class1})-(\ref{class3}) have to satisfy the classical CH inequality (\ref{CH}). Since for appropriate choice of the measurement directions this inequalitity is violated, EPR correlations cannot be given a \textit{classical} local, non-conspiratorial joint common causal explanation.
\vspace{0.1in}

\noindent
Observe that up to this point everything has been classical. Quantum mechanics (QM) was simply used to generate classical conditional probabilities by the Born rule. These conditional probabilities, however, could also have been directly obtained from the laboratory and in the actual experiments they are gained in this direct way indeed. So it is completely satisfactory to interpret the EPR scenario---in accord with the quote from Gisin in the Introduction---as a classical situation with classical conditional correlation (between detector clicks) violating the classical CH inequality (\ref{CH}) (see (Szab\'o 1998)).

But this is \textit{not} the standard interpretation. The standard way to describe the above EPR situation is to adopt another mathematical formalism, the formalism of quantum theory. Here events are represented as projections of the von Neumann lattice of the tensor product matrix algebra $M_2(\CO)\otimes M_2(\CO)$ and probabilities are gained by the quantum states. So instead of (\ref{class1})-(\ref{class3}) one writes the following:
\begin{eqnarray}
\phi^s(A_m) &=& Tr\big(\rho^s \, (A_m \otimes \UN_B)\big) = \frac{1}{2} \label{quant1} \\
\phi^s(B_n) &=& Tr\big(\rho^s \, (\UN_A \otimes B_n)\big) = \frac{1}{2}\label{quant2} \\
\phi^s(A_m B_n) &=& Tr\big(\rho^s \,( A_m \otimes B_n)\big) = \frac{1}{2} \sin^2\left(\frac{\theta _{a_m b_n}}{2}\right) \label{quant3}
\end{eqnarray}
where $A_m$ and $B_n$ denote projections onto the eigensubspaces with eigenvalue $+\frac{1}{2}$ of the spin operators associated with directions $\vec{a}_{m}$ and $\vec{b}_{n}$, respectively, and $\phi^s( \, \cdot \, ) = Tr(\rho^s \, \cdot \, )$ is the singlet state. Moreover, if we go over to AQFT, these projections will be localized in a well-defined spacetime region.

Substituting the \textit{non-classical} probabilities (\ref{quant1})-(\ref{quant3}) into the \textit{non-classical} CH inequality (\ref{qCH}) defined in the Introduction one finds a violation of this inequality for appropriate choices of the projections $A_m, B_n$. But what does it mean? First, it is important to be aware of the fact that now we adopt another theory to account for correlations. But then we need to take the consequences of this move seriously. This means that we need to represent \textit{every} event of the model as projections of a von Neumann algebra. Among them common causes! So the following questions arise: Can the classical notion of the common cause (system) generalized for the non-classical case? What is the relation of this non-classical notion of common cause to the non-classical CH inequality (\ref{qCH})? Does there exist a non-classical common causal justification of the Bell inequalities used in AQFT similar to the classical one?

As it will turn out soon, one can generalize the notion of the common cause also for the algebraic quantum field theoretical setting, and one can also give a precise definition of a local, non-conspiratorial joint common causal explanation of a set of correlations in AQFT. However, it also will turn out that there is no direct relation between this common causal explanation and the Bell inequalities. Or to put it briefly, correlation violating the Bell inequality can still have a local, non-conspiratorial joint common causal explanation. In order to see all these, first we have to generalize the notions of this Section to the quantum case.

\section{Non-classical common causal explanation}\label{non-class}

Let us first generalize the notion of the common cause system to the quantum case in the following way. Replace the classical probability measure space $(\Omega, \Sigma,p)$ by the non-classical probability measure space $(\mathcal{N}, \PR(\mathcal{N}), \phi)$ where $\PR(\vNA)$ is the (non-distributive) lattice of projections (events) and $\phi$ is a state of a von Neumann algebra $\vNA$. We note that in case of projection lattices we will use only algebra operations (products, linear combinations) instead of lattice operations ($\vee,\wedge$). In case of commuting projections $A,B\in\PR(\vNA)$ lattice operations can be given in terms of algebraic operations.

A set of mutually orthogonal projections $\left\{ C_k \right\}_{k\in K}\subset\mathcal{P}(\vNA)$ is called a \textit{partition of the unit} $\UN\in\vNA$ if $\sum_k C_k = \UN$. Two commuting projections $A$ and $B \in \mathcal{P}(\mathcal{N})$ are said to be correlating in the state $\phi\colon\mathcal{N}\to\CO$ if
\begin{eqnarray}\label{qcorr}
\phi(AB) & \neq & \phi(A)\, \phi(B).
\end{eqnarray}
Since $\phi$ is linear, a kind of `theorem of total probablity',  $\sum_i\phi(AP_i)=\phi(A\sum_iP_i)=\phi(A)$, holds for any partition $\left\{ P_i \right\}$ of the unit, hence (\ref{qcorr}) is equivalent to
\begin{eqnarray}\label{qcorrrew}
\phi(AB)\, \phi(A^{\perp}B^{\perp}) & \neq & \phi(AB^{\perp})\, \phi(A^{\perp}B).
\end{eqnarray}
Now, following the lines of Definition \ref{ccs} one can characterize the non-classical common cause system of the correlation (\ref{qcorr}) as a screener-off partition of the unit. To make the definition meaningful we have to introduce the following \textit{conditional expectation} $E_c\colon\vNA\to{\cal{C}}$:
\begin{equation}\label{ncqcorr}
E_c(A):=\sum_{k\in K} C_kAC_k,
\end{equation}
where $\{ C_k \}_{k\in K}$ is a partition of the unit of $\vNA$ (Umegaki, 1954). The image {\cal{C}} of this map is a unital subalgebra of $\vNA$ containing exactly those elements that commute with $C_k,k\in K$. Therefore, $E_c(A)C_k=E_c(AC_k)=C_kAC_k$ ($A\in\vNA, k\in K$) for example. By means of this conditional expectation we can define the notion of the common cause system in the non-classical case:
\begin{D}\label{ncqccs}
A partition of the unit $\left\{ C_k \right\}_{k\in K}\subset\mathcal{P}(\mathcal{N})$ is said to be the {\em common cause system} of the commuting events $A,B\in\cal{P}(\vNA)$, which correlate in the state $\phi\colon\vNA\to\CO$, if for those $k\in K$ for which $\phi(C_k)\not= 0$, the following condition holds:
\begin{eqnarray}\label{ncqccs0}
\frac{(\phi\circ E_c)(ABC_k)}{\phi(C_k)}&=& \frac{(\phi\circ E_c)(AC_k)}{\phi(C_k)} \frac{(\phi\circ E_c)(BC_k)}{\phi(C_k)}\label{ncqccs1} . 
\end{eqnarray}
If $C_k$ commutes with both $A$ and $B$ for all $k\in K$, we call $\left\{ C_k \right\}_{k\in K}$ a {\em commuting} common cause system, otherwise a {\em noncommuting} one. A common cause system of size $\vert K\vert=2$ is called a {\em common cause}.
\end{D}
Some remarks are in place here. First, using the `theorem of total probability' the common cause condition (\ref{ncqccs0}) can be written as
\begin{eqnarray}\label{ncqccs2}
(\phi\circ E_c)(ABC_k))\, (\phi\circ E_c)(A^{\perp}B^{\perp}C_k) &=& (\phi\circ E_c)(AB^{\perp}C_k)\,(\phi\circ E_c) (A^{\perp}BC_k),\ k\in K.
\end{eqnarray} 
One can even allow here the case $\phi(C_k)=0$, since then both sides of (\ref{ncqccs2}) are zero.

Second, the non-classical character of the common cause system of Definition \ref{ncqccs} lies in the fact that the common cause system need \textit{not} commute with the correlating events. If the events $A$ and $B$ commute with $C_k, k\in K$, then not only $C_k\in{\cal{C}}$ but also $A,B, A^\perp, B^\perp\in{\cal{C}}$, and therefore $E_c(ABC_k)=ABC_k$, for example. Thus, the conditional expectation $E_c$ vanishes from the defining equation (\ref{ncqccs0}); and (\ref{ncqccs2}) leads to
\begin{eqnarray}\label{qccs1} 
\phi(ABC_k)\, \phi(A^{\perp}B^{\perp}C_k)&=&\phi(AB^{\perp}C_k)\, \phi(A^{\perp}BC_k).
\end{eqnarray}

Finally, it is obvious from (\ref{qccs1}) that if $C_k\leq X$ with $X=A,A^\perp, B$ or $B^\perp$ for any $k\in K$ then $\left\{ C_k \right\}_{k\in K}$ serve as a common cause system (and hence a commuting common cause system) of the given correlation independently of the chosen state $\phi$. These solutions are called \textit{trivial common cause systems}. In case of common cause, $\vert K\vert=2$, triviality means that $\{C_k\}=\{ A,A^\perp\}$ or $\{C_k\}=\{ B,B^\perp\}$. 

Having generalized the notion of the common cause system for the quantum case, the next step is to localize it. Suppose that the projection $A$ is localized in the algebra $\mathcal A(V_A)$ with support $V_A$ and the projection $B$ is localized in the algebra $\mathcal A(V_B)$ with support $V_B$ such that $V''_A$ and $V''_B$ are spacelike separated double cones in a spacetime $\mathcal S$. A common cause system $\{ C_k \}_{k\in K}$ is said to be a \textit{commuting/noncommuting (strong/weak) common cause system }of the correlation between $A$ and $B$ if $\{ C_k \}_{k\in K}$ is localizable in an algebra  $\mathcal A(V_C)$ with support $V_C$ such that $V_C$ is in $cpast(V_A, V_B)$ ($spast(V_A, V_B)$/$wpast(V_A, V_B)$).

In the same vein, we obtain the definition of the \textit{joint common cause system} in the non-classical case. Let $\{(A_m, B_n); m \in M,n \in N\}$ be a set of pairs of commuting projections correlating in the sense that
\begin{eqnarray} \label{qcorrs}
\phi(A_mB_n) &\neq& \phi(A_m)\, \phi(B_n).
\end{eqnarray}
\begin{D}\label{qcccs} 
A partition of the unit $\left\{ C_k \right\}_{k\in K}\subset\mathcal{P}(\mathcal{N})$ is said to be a {\em joint common cause system} of the set $\{(A_m, B_n); m \in M,n \in N\}$ of commuting pairs of correlating events, if for any $k\in K$, when $\phi(C_k)\not= 0$, the conditions 
\begin{eqnarray}\label{qcccs1}
\frac{(\phi\circ E_c)(A_mB_nC_k)}{\phi(C_k)}&=& \frac{(\phi\circ E_c)(A_mC_k)}{\phi(C_k)} \frac{(\phi\circ E_c)(B_nC_k)}{\phi(C_k)},\quad m\in M, n\in N\label{ncqccs1} 
\end{eqnarray}
hold, where $E_c$ is the conditional expectation defined in (\ref{ncqcorr}). Again, if $\left\{ C_k \right\}_{k\in K}$ commutes with $A_m$ and $B_n$ for all $m\in M, n\in N$, then we call it a {\em commuting} joint common cause system, otherwise a {\em noncommuting} one. 
\end{D}
Equation (\ref{qcccs1}) can again be understood in the more permissive way as
\begin{eqnarray}\label{qcccs2}
(\phi\circ E_c)(A_m B_n C_k))\, (\phi\circ E_c)(A^{\perp}_m B^{\perp}_n C_k) &=& (\phi\circ E_c)(A_m B^{\perp}_n C_k)\, (\phi\circ E_c) (A^{\perp}_m B_n C_k)
\end{eqnarray}
incorporating cases when $\phi(C_k)= 0$.

And here comes a subtle point. Having introduced the notion of the joint common cause system of a correlation in the preceding Section we went over to conditional correlations and defined a \textit{local, non-conspriratorial} common causal explanation of these correlations. What is the analogue move in the non-classical case? We claim that we need \textit{not} introduce any new concept; the definition of a \textit{local, non-conspriratorial} common cause system in the non-classical case is just \textit{identical} to the one given in Definition \ref{qcccs} that is to the definition of the joint common cause system. For the details see the Appendix (and (Butterfield 1995)). So from now on we drop the prefix `local, non-conspiratorial' before the term `joint common cause system' in the non-classical case.
\vspace{0.1in}

\noindent
Now, we are able to ask whether there is a proposition similary to Proposition \ref{PCH} in the non-classical case, that is whether one can derive a CH inequality (\ref{qCH}) from the fact that the set of correlating projections $\{(A_m, B_n); m \in M,n \in N\}$ has a \textit{joint common causal explanation}? The following proposition provides a sufficient condition.

\begin{Prop}\label{PqCH} 
Let $A_m  \in \mathcal A(V_A)$ and  $B_n \in \mathcal A(V_B)$  $(m,n=1,2)$ be four projections localized in spacelike separated spacetime regions $V_A$ and $V_B$, respectively, which correlate in the locally faithful state $\phi$ in the sense of (\ref{qcorrs}). Suppose that $\{(A_m, B_n); m,n=1,2\}$ has a joint common causal explanation in the sense of Definition \ref{qcccs}. Then for any $m,m', n, n' = 1,2; m\neq m'; n\neq n'$ the CH inequality 
\begin{eqnarray}\label{qCH_mn_noncomm} 
-1 \leqslant (\phi\circ E_c) (A_m B_n + A_m B_{n'} + A_{m'} B_n - A_{m'} B_{n'} - A_m - B_n ) \leqslant 0.
\end{eqnarray}
holds for the state $\phi\circ E_c$. If the joint common cause is a \textit{commuting} one, then the CH inequality holds for the original state $\phi$: 
\begin{eqnarray}\label{qCH_mn} 
-1 \leqslant \phi (A_m B_n + A_m B_{n'} + A_{m'} B_n - A_{m'} B_{n'} - A_m - B_n ) \leqslant 0.
\end{eqnarray}
\end{Prop}

\noindent\textit{Proof.} Substituting the expressions
\begin{eqnarray}
\alpha  &:=& \frac{(\phi\circ E_c)(A_mC_k)}{\phi(C_k)} \label{qa} \\
\alpha ' &:=& \frac{(\phi\circ E_c)(A_{m'}C_k)}{\phi(C_k)}  \label{aa'} \\
\beta  &:=& \frac{(\phi\circ E_c)(B_nC_k)}{\phi(C_k)} \label{qb} \\
\beta ' &:=& \frac{(\phi\circ E_c)(B_{n'}C_k)}{\phi(C_k)} \label{qb'}
\end{eqnarray}
into the inequality
\begin{eqnarray*}
-1 \leqslant \alpha \beta +\alpha \beta '+\alpha '\beta -\alpha '\beta '-\alpha -\beta \leqslant 0
\end{eqnarray*}
and using (\ref{qcccs1}) we get
\begin{eqnarray}
-1 \leqslant \frac{(\phi\circ E_c)(A_mB_nC_k)}{\phi(C_k)} + \frac{(\phi\circ E_c)(A_mB_{n'}C_k)}{\phi(C_k)} + \frac{(\phi\circ E_c)(A_{m'}B_nC_k)}{\phi(C_k)} \nonumber \\
- \frac{(\phi\circ E_c)(A_{m'}B_{n'}C_k)}{\phi(C_k)} - \frac{(\phi\circ E_c)(A_mC_k)}{\phi(C_k)} - \frac{(\phi\circ E_c) (B_nC_k)}{\phi(C_k)} \leqslant 0 .
\end{eqnarray}  
Multiplying the above inequality by $\phi(C_k)$ and summing up for the index $k$ one obtains
\begin{eqnarray}\label{qCH_C} 
-1 \leqslant \sum_k \bigg( (\phi\circ E_c)(A_mB_nC_k) + (\phi\circ E_c)(A_mB_{n'}C_k) + (\phi\circ E_c)(A_{m'}B_nC_k) \nonumber \\
- (\phi\circ E_c)(A_{m'}B_{n'}C_k) - (\phi\circ E_c)(A_mC_k) - (\phi\circ E_c) (B_nC_k) \bigg) \leqslant 0,
\end{eqnarray}
which leads to (\ref{qCH_mn_noncomm}) by performing the summation. If $\left\{ C_k \right\}_{k\in K}$ is a \textit{commuting} joint common cause system, then $E_c$ drops out from the above expression since all the arguments are in ${\cal{C}}$ (see the remark before (\ref{ncqccs2})). Therefore (\ref{qCH_C}) becomes identical to (\ref{qCH_mn}), which completes the proof.\qed 
\vspace{0.1in}

\noindent
First note that similarly to Proposition \ref{PCH}, neither Proposition \ref{PqCH} refers to the spacetime localization of $\{ C_k\}$ in a direct way. Indirectly, however, it restricts the localization of the possible joint common cause systems for states violating the CH inequality (\ref{qCH_mn}): the support of $\{ C_k\}$ \textit{must} intersect the union of the causal past or the causal future of $V_A\cup V_B$. It is so because otherwise the support of $\left\{ C_k \right\}_{k\in K}$ would be spacelike separated from those of $A$ and $B$, and hence $\{ C_k\}$ would be a \textit{commuting} joint common cause system for a set of correlations violating the CH inequality (\ref{qCH_mn}), in contradiction with Proposition \ref{PqCH}.

Proposition \ref{PqCH}---similarly to Proposition \ref{PCH}---provides a \textit{common causal justification} of the CH inequality (\ref{qCH_mn}). It states that in order to yield a \textit{commuting} joint common causal explanation for the set $\{(A_m, B_n); m,n=1,2\}$ the CH inequality (\ref{qCH_mn}) has to be satisfied. But what is the situation with \textit{noncommuting} common cause systems? Since---apart from (\ref{qCH_mn_noncomm})---Proposition \ref{PqCH} is silent about the relation between a \textit{noncommuting} joint common causal explanation and the CH inequality (\ref{qCH_mn}), the question arises: Can a \textit{set} of correlations violating the CH inequality (\ref{qCH_mn}) have a \textit{noncommuting} joint common causal explanation? Before addressing this question, we pose an easier one: Can \textit{a single} correlation have a common causal explanation in AQFT? This leads us over to the question of the validity of the Common Cause Principles in AQFT.

\section{Common Cause Principles in algebraic quantum field theory}

Reichenbach's Common Cause Principle (CCP) is the following hypothesis: If there is a correlation between two events and there is no direct causal (or logical) connection between the correlating events, then there exists a common cause of the correlation. The precise definition of this informal statement that fits to the algebraic quantum field theoretical setting is the following:
\begin{D}\label{CCP}
A $\mathcal{P}_\SDK$-covariant local quantum theory $\{\OA(V),V\in\SDK\}$ is
said to satisfy the Commu\-tative/Noncommutative (Weak/Strong) Common Cause
Principle if for any pair $A \in\mathcal A(V_1)$ and $B\in\mathcal A(V_2)$ of
projections supported in spacelike separated regions $V_1, V_2\in\SDK$ and
for every locally faithful state $\phi\colon\OA\to\CO$ establishing a correlation between $A$ and $B$, there exists a \textit{nontrivial} commuting/noncommuting common cause system
$\{ C_k \}_{k\in K}\subset \mathcal A(V), V\in\SDK$ of the correlation
(\ref{qcorr}) such that the localization region $V$ is in the (weak/strong) common past of $V_1$ and $V_2$. 
\end{D}
What is the status of these six different notions of the Common Cause Principle in AQFT?

The question whether the Commutative Common Cause Principles are valid in a
Poincar\'e covariant local quantum theory in the von Neumann algebraic setting was first raised by R\'edei (1997, 1998). As an answer to this question, R\'edei and
Summers (2002, 2007) have shown that the Commutative Weak CCP is valid in
algebraic quantum field theory with locally infinite degrees of freedom. Namely,
in the von Neumann setting they proved that for every locally normal and
faithful state and for every superluminally correlating pair of projections
there exists a weak common cause, that is a common cause system of size 2 in
the weak past of the correlating projections. They have also shown (R\'edei and
Summers, 2002, p 352) that the localization of a common cause $C<AB$ cannot be restricted to $wpast(V_1, V_2) \setminus I_-(V_1)$ or $wpast(V_1, V_2) \setminus I_-(V_2)$ due to logical independence of spacelike separated algebras.

Concerning the Commutative (Strong) CCP less is known. If one also admits
projections localized only in \textit{un}bounded regions, then the Strong CCP is
known to be false: von Neumann algebras pertaining to complementary wedges
contain correlated projections but the strong past of such wedges is empty (see
(Summers and Werner, 1988) and (Summers, 1990)). In spacetimes having horizons, 
e.g. those with Robertson--Walker metric, the common past of spacelike separated bounded regions can be empty, although there are states which provide correlations among local algebras corresponding to these regions (Wald 1992).\footnote{We thank David Malament for calling our attention to this point and the paper of Wald.} Hence, CCP is not valid there. Restricting ourselves to \textit{local} algebras in Minkowski spaces the situation is not clear. We are of the opinion that one cannot decide on the validity of the (Strong) CCP without an explicit reference to the dynamics since there is no bounded region $V$ in $cpast(V_1, V_2)$ (hence
neither in $spast(V_1, V_2)$) for which isotony would ensure that $\OA(V_1\cup
V_2)\subset\OA(V'')$. But dynamics relates the local algebras since $\OA(V_1\cup
V_2)\subset\OA(V''+t) =\alpha_t(\OA(V''))$ can be fulfilled for certain $V\subseteq V''\subset cpast(V_1, V_2)$ and for certain time translation by $t$. 

Coming back to the proof of R\'edei and Summers, the proof had a crucial premise,
namely that the algebras in question are \textit{von Neumann algebras of type
III}. Although these algebras arise in a natural way in the context of
Poincar\'e covariant theories, other local quantum theories apply von Neumann
algebras of other type. For example, theories with locally finite degrees of
freedom are based on finite dimensional (type I) local von Neumann algebras.
This raised the question whether the Commutative Weak CCP is valid in other
local quantum theories. To address the problem Hofer-Szab\'o and Vecserny\'es (2012a) have chosen the local quantum Ising model (see M\"uller, Vecserny\'es) having locally finite degrees of freedom. It turned out that the Commutative Weak CCP is \textit{not valid} in the local quantum Ising model and it cannot be valid either in theories with locally finite degrees of freedom in general.

But why should we require commutativity between the common cause and its effects at all?

Commutativity has a well-defined role in any quantum theories: observables should commute to be simultaneously measurable. In AQFT commutativity of observables with spacelike separated supports is an axiom. To put it simply, commutativity can be required for events which can happen `at the same time'. But cause and effect are typically \textit{not} this sort of events. If one considers ordinary QM, one well sees that observables do not commute even with their own time translates in general. For example, the time translate $x(t) := U(t)^{-1}xU(t)$ of the position operator $x$ of the harmonic oscillator in QM does \textit{not} commute with $x\equiv x(0)$ for generic $t$, since in the ground state vector $\psi_0$ we have
\begin{eqnarray}
\big[ x, x(t)\big] \, \psi_0 = \frac{-i\hbar \sin{(\hbar\omega t)}}{m\omega} \psi_0 \not\equiv 0.
\end{eqnarray}
Thus, if an observable $A$ is not a conserved quantity, that is $A(t)\not= A$,
then the commutator $[A,A(t)]\not= 0$ in general. So why should the commutators
$[A,C]$ and $[B,C]$ vanish for the events $A,B$ and for their common cause
$C$ supported in their (weak/common/strong) past? We think that commuting common causes are only unnecessary reminiscense of their classical formulation. Due to their relative spacetime localization, that is due to the time delay between the correlating events and the common cause, it is also an unreasonable assumption.

Abandoning commutativity in the definition of the common cause is therefore a natural move. To our knowledge the first to contemplate the possibility of the noncommuting common causes were Clifton and Ruetsche (1999) in their paper criticizing R\'edei (1997, 1998) who required commutativity from the common cause. They say: ``[requiring commutativity] bars form candidacy to the post of common cause the vast majority of events in the common past of events problematically correlated'' (p 165). And indeed, the benefit of allowing noncommuting common causes is that the noncommutative version of the result of R\'edei and Summers can be regained: as it was shown in (Hofer-Szab\'o and Vecserny\'es 2012b), by allowing common causes that do \textit{not} commute with the correlating events, the Weak CCP can be proven in local UHF-type quantum theories.

Now, let us turn to our original question as to whether a \textit{set} of correlations violating the CH inequality (\ref{qCH}) can have a noncommuting \textit{joint} common causal explanation in AQFT. Since our answer is provided in  an AQFT with locally finite degrees of freedom, in the local quantum Ising model, we give a short and non-technical tutorial to this model in the next Section. (For more detail see (Hofer-Szab\'o, Vecserny\'es, 2012c).) 

\section{Noncommutative common causes for correlations violating the CH inequality}

\begin{figure}[ht]
\centerline{\resizebox{6cm}{!}{\includegraphics{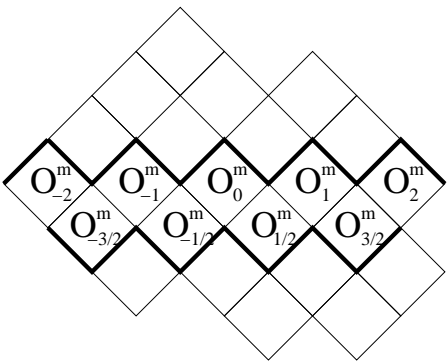}}}
\caption{A thickened Cauchy surface in the two dimensional Minkowski space $\Mink^2$}
\label{strong_Bell1}
\end{figure}

Consider a `discretized' version of the two dimensional Minkowski spacetime $\mathcal M^2$ which is composed of minimal double cones $\DK^m(t,i)$ of unit diameter with their center in $(t,i)$ for $t,i\in\mathbb{Z}$ or $t,i\in\mathbb{Z}+1/2$. The set $\{\DK^m_i,  i\in\fel\mathbb{Z}\}$  of such minimal double cones with $t=0, -1/2$ defines a `thickened' Cauchy surface in this spacetime (see Fig. \ref{strong_Bell1}). The double cone $\DK^m_{i,j}$ sticked to this Cauchy surface is defined to be the smallest double cone containing both $\DK_i^m$ and $\DK_j^m$: $\DK^m_{i,j}:=\DK_i^m\vee \DK_j^m$. Similarly, let $\DK^m(t,i;s,j):=\DK^m(t,i)\vee \DK^m(s,j)$. The directed set of such double cones is denoted by $\SDK^m$, and the directed subset of it whose elements are sticked to a Cauchy surface is denoted by $\SDK^m_{CS}$. Obviously, $\SDK^m_{CS}$ will be left invariant by integer space translations and $\SDK^m$ will be left invariant by integer space and time translations.

The net of local algebras is defined as follows. The `one-point' observable algebras associated to the minimal double cones $\DK^m_i,i\in\fel\mathbb{Z}$ are defined to be $\OA(\DK^m_i)\simeq M_1(\CO)\oplus M_1(\CO)$. Between the unitary selfadjoint generators $U_i\in\OA(\DK^m_i)$ one demands the following commutation relations:
\begin{eqnarray}\label{comm_rel}
U_i U_j = \left\{ \begin{array}{rl} -U_j U_i, & \mbox{if}\ |i-j|=\frac{1}{2},
\\ U_j U_i, & \mbox{otherwise.}\ \end{array} \right.
\end{eqnarray}
Now, the local algebras $\OA(\DK_{i,j}), \DK_{i,j}\in\SDK^m_{CS}$ are linearly spanned by the monoms
\begin{eqnarray}\label{monoms}
U_i^{k_i} \, U_{i+\frac{1}{2}}^{k_{i+\frac{1}{2}}} \,  \dots \, U_{j-\frac{1}{2}}^{k_{j-\frac{1}{2}}} \, U_j^{k_j} 
\end{eqnarray}
where $k_i, k_{i+\frac{1}{2}} \dots k_{j-\frac{1}{2}}, k_j \in \{0,1\}$.\footnote{For detailed Hopf algebraic description of the local quantum spin models see (Szlach\'anyi, Vecserny\'es, 1993), (Nill, Szlach\'anyi, 1997), (M\"uller, Vecserny\'es)).} 

Since the local algebras $\OA(\DK_{i,i-\frac{1}{2}+n}), \, i\in\frac{1}{2}\mathbb{Z}$ for $n\in \mathbb{N}$ are isomorphic to the full matrix algebra $M_{2^n}(\mathbb{C})$, the quasilocal observable algebra $\OA$ is a uniformly hyperfinite (UHF) $C^*$-algebra and consequently there exists a unique (non-degenerate) normalized trace $\textrm{Tr}\colon\OA\to\CO$ on it. We note that all nontrivial monoms in (\ref{monoms}) have zero trace. 

\begin{figure}[ht]
\centerline{\resizebox{6cm}{!}{\includegraphics{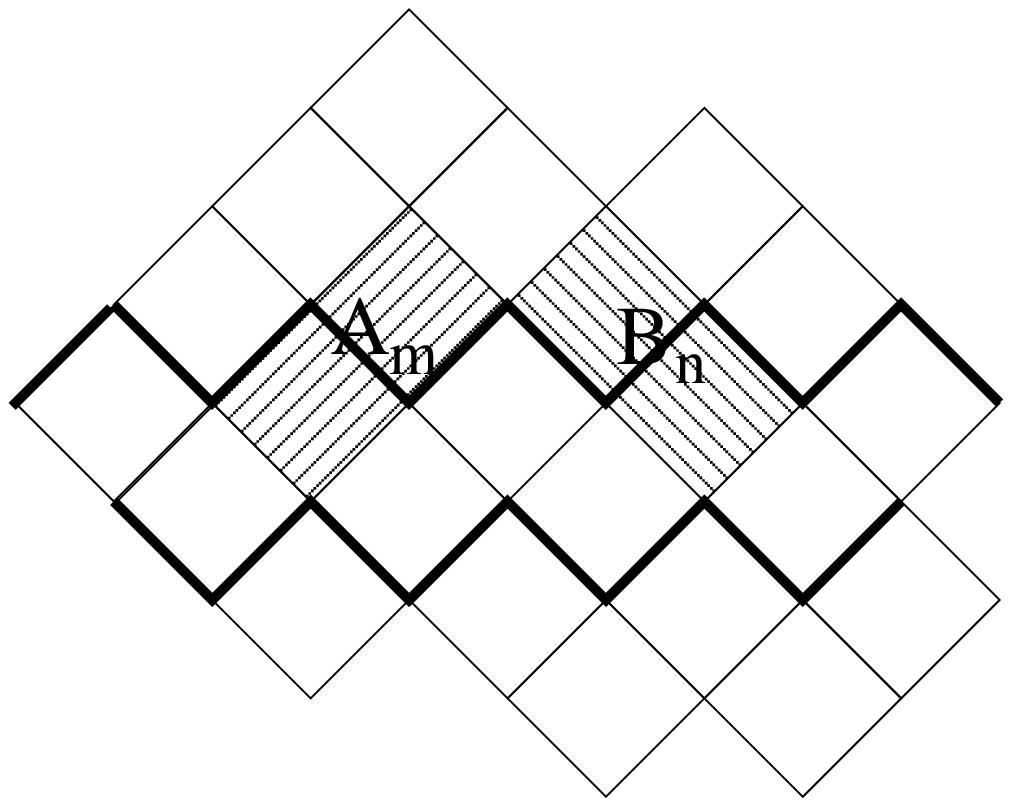}}}
\caption{Projections in $\OA(\DK_A)$ and $\OA(\DK_B)$}
\label{strong_Bell2}
\end{figure}

In order to extend the `Cauchy surface net' $\{\OA(\DK),\DK\in\SDK^m_{CS}\}$ to the net $\{\OA(\DK),\DK\in\SDK^m\}$ in a causal and time translation covariant manner one has to classify causal (integer valued) time evolutions in the local quantum Ising model. This classification was given in (M\"uller, Vecserny\'es) and it also was shown that the extended net satisfies isotony, Einstein causality, algebraic Haag duality
\begin{eqnarray}\label{Haag} 
\OA(\DK')'\cap\OA =\OA(\DK),\quad\DK\in\SDK^m,
\end{eqnarray}
$\IN\times\IN$ covariance with respect to integer time and space translations and primitive causality:
\begin{eqnarray}\label{primcaus} 
\OA(V)=\OA(V''),
\end{eqnarray}
where $V$ is a finite connected piece of a thickened Cauchy surface (composed of minimal double cones). $V''$ denotes the double spacelike complement of $V$, which is the smallest double cone in $\SDK^m$ containing $V$. We will be interested here only in a special subset of these causal automorphisms given by:
\begin{eqnarray}\label{causal_automorph}
\beta(U_x)&=& U_{x-\fel}U_xU_{x+\fel}, \quad x\in\IN+\fel.
\end{eqnarray}
(In our following example we need not specify the choice for $\beta(U_x), x\in\IN$.) Now, consider the double cones $\DK_A:=\DK^m(0,-1) \, \cup \, \DK^m(\fel,-\fel)$ and $\DK_B:=\DK^m(\fel,\fel) \, \cup \, \DK^m(0,1)$ and the `two-point' algebras $\OA(\DK_A)$ and $\OA(\DK_B)$ pertaining to them. (See Fig. \ref{strong_Bell2}.) 
A linear basis of the algebra $\OA(\DK_A)$ is given by the monoms 
\begin{eqnarray}\label{monoms2}
\UN,\qquad U_{-1}, \qquad \beta(U_{-\fel}) \equiv U_{-1} U_{-\fel} U_0,\qquad iU_{-1}\beta(U_{-\fel}) \equiv iU_{-\fel} U_0
\end{eqnarray}
(where $i$ in the fourth monom is the imaginary unit). They satisfy the same commutation relations like the Pauli matrices $\sigma_0 = \UN, \sigma_x, \sigma_y$ and $\sigma_z$ in $M_2(\CO)$. Therefore, introducing the notation 
\begin{eqnarray}
{\bf U} &:=& (U_{-1}, \, U_{-1} U_{-\fel} U_0, \, iU_{-\fel} U_0)
\end{eqnarray}
any minimal projection in $\OA(\DK_A)$ can be parametrized as 
\begin{eqnarray}
A({\bf a})&:=&\fel\left( \UN+ {\bf a} {\bf U} \right)  \label{A_m}
\end{eqnarray}
where ${\bf a} = (a_1, a_2, a_3)$ is a unit vector in $\RE^3$. In the same vein, any minimal projection in $\OA(\DK_B)$ can be paremetrized as 
\begin{eqnarray}
B({\bf b})&:=&\fel\left( \UN+ {\bf b} {\bf V} \right)  \label{B_n} 
\end{eqnarray}
where 
\begin{eqnarray}
{\bf V} &:=& (U_1, \, -U_0 U_{\fel} U_{1}, \, iU_0 U_{\fel})
\end{eqnarray} 
is the vector composed of the generators of $\OA(\DK_B)$ and ${\bf b} = (b_1, b_2, b_3)$ is a unit vector in $\RE^3$. The projections $A({\bf a})$ and $B({\bf b})$ can be interpreted as the event localized in $\OA(\DK_A)$ and $\OA(\DK_B)$, respectively pertaining to the generalized spin measurement in direction ${\bf a}$ and ${\bf b}$, respectively.

Now, consider two projections $A_m:= A({\bf a^m});m=1,2$ localized in $\DK_A$, and two other projections $B_n:= B({\bf b^n});n=1,2$ localized in the spacelike separated double cone $\DK_B$. Suppose that our system is in the faithful state $\phi(\, \cdot \,) = Tr(\rho\,\cdot\,)$
where
\begin{eqnarray} \label{rho} 
\rho &=&\rho(\lambda):=  \UN + \lambda \big( U_{-1} U_{-\frac{1}{2}} U_{\frac{1}{2}} U_1 - U_{-1} U_1 + U_{-\frac{1}{2}} U_{\frac{1}{2}} \big),\quad \lambda\in [0,1).
\end{eqnarray}
For $\lambda = 1$ the state defined by (\ref{rho}) gives us back the usual singlet state. It is easy to see that in the state (\ref{rho}) the correlation between $A_m$ and $B_n$ will be:
\begin{eqnarray}\label{corr} 
corr(A_m,B_n) := \phi(A_m B_n) - \phi(A_m)\, \phi(B_n) = - \frac{\lambda}{4} \left\langle {\bf a^m}, {\bf b^n}\right\rangle 
\end{eqnarray}
where $\left\langle \, \, , \, \, \right\rangle$ is the scalar product in $\RE^3$. In other words $A_m$ and $B_n$ will correlate whenever ${\bf a^m}$ and ${\bf b^n}$ are not orthogonal. Now, if ${\bf a^m}$ and ${\bf b^n}$ are chosen as
\begin{eqnarray}
{\bf a^1} &=& (0,1,0) \label{a1} \\
{\bf a^2} &=& (1,0,0) \\
{\bf b^1} &=& \frac{1}{\sqrt{2}} (1,1,0) \\
{\bf b^2} &=& \frac{1}{\sqrt{2}} (-1,1,0) \label{b2} 
\end{eqnarray}
the CH inequality (\ref{qCH}) will be violated at the lower bound since
\begin{eqnarray}\label{CH'} 
\phi( A_1 B_1 + A_1 B_2 + A_2 B_1 - A_2 B_2 - A_1 - B_1 \big) = \nonumber \\
- \frac{1}{2} - \frac{\lambda}{4} \left(  \left\langle {\bf a^1}, {\bf b^1} \right\rangle + \left\langle {\bf a^1}, {\bf b^2} \right\rangle + \left\langle {\bf a^2}, {\bf b^1} \right\rangle - \left\langle {\bf a^2}, {\bf b^2} \right\rangle \right) = - \frac{1+\lambda\sqrt{2}}{2},
\end{eqnarray}
which is smaller than $-1$ if $\lambda > \frac{1}{\sqrt{2}}$. Or, equivalently, the CHSH inequality (\ref{qCHSH}) where
\begin{eqnarray}\label{unitaries} 
X_m &:=& 2A_m - \UN \\
Y_n &:=& 2B_n - \UN
\end{eqnarray}
will be violated for the above setting since
\begin{eqnarray}\label{CHSH'} 
\phi(X_1(Y_1 + Y_2) + X_1(Y_1 - Y_2)) = \nonumber \\
=  - \lambda \left( \left\langle {\bf a^1}, {\bf b^1} + {\bf b^2}\right\rangle + \left\langle {\bf a^2}, {\bf b^1} - {\bf b^2}\right\rangle  \right) = -\lambda2\sqrt{2} 
\end{eqnarray}
is smaller than $-2$ if $\lambda > \frac{1}{\sqrt{2}}$. Both the CH and the CHSH inequality are maximally violated for the singlet state, that is if $\lambda = 1$.

\begin{figure}[ht]
\centerline{\resizebox{5cm}{!}{\includegraphics{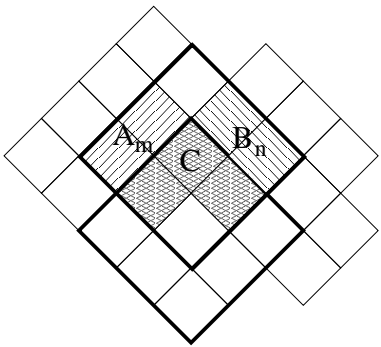}}}
\caption{Localization of a common cause for the correlations $\{(A_m, B_n)\}$.}
\label{strong_Bell4}
\end{figure}

The  question whether the four correlations $\{(A_m, B_n); m,n=1,2\}$ violating the CH inequality (\ref{qCH}) have a joint common causal explanation was answered in (Hofer-Szab\'o, Vecserny\'es, 2012c) by the following 
\begin{Prop}\label{NCCCCS}
Let $A_m:=A({\bf a^m})\in\OA(\DK_A), B_n:= B({\bf b^n})\in\OA(\DK_B); m,n=1,2$ be four projections defined in (\ref{A_m})-(\ref{B_n}), where ${\bf a^m}$ and ${\bf b^n}$ are non-orthogonal unit vectors in $\RE^3$ establishing four correlations $\{(A_m, B_n); m,n=1,2\}$ in the state (\ref{rho}). Let furthermore $C$ be any projection localized in $\DK_C:= \DK_{-\fel}\vee\DK_\fel\in\SDK^m_{CS}$ (see Fig. \ref{strong_Bell4}.) of the shape
\begin{eqnarray}\label{C} 
C &=& \frac{1}{4}\left(\UN + U_{-\fel} U_{\fel}\right)\left(\UN+c_1 U_0+c_2 U_{\fel} + c_3 iU_0 U_{\fel}\right) \nonumber \\
& & + \frac{1}{4}\left(\UN - U_{-\fel} U_{\fel}\right)\left(\UN+c'_1 U_0+c'_2 U_{\fel} + c'_3 iU_0 U_{\fel}\right)
\end{eqnarray}
where ${\bf c} = (c_1, c_2, c_3)$ and ${\bf c'} = (c'_1, c'_2, c'_3)$ are arbitrary unit vectors in $\RE^3$. Then $\{C, C^{\perp}\}$ is a joint common cause of the correlations $\{(A_m, B_n)\}$ if $a^m_3b^n_3 = 0$ for any $m,n=1,2$ and $c_2 =0$.
\end{Prop}

Since for the directions ${\bf a^m}$ and ${\bf b^n}$ defined in (\ref{a1})-(\ref{b2}) the requirement $a^m_3b^n_3 = 0$ holds for any $m,n=1,2$, therefore the correlations (maximally) violating the CH/CHSH inequality \textit{do} have a joint common cause---any $C$ of form (\ref{C}) with $c_2 =0$. 

Finally, here is a Proposition (consistently with the derivability of a CH inequality from the commuting joint common cause system) claiming that there exists no \textit{commuting} joint common cause for these correlations even without any restriction to their localization (Hofer-Szab\'o, Vecserny\'es, 2012c):
\begin{Prop}\label{nocommjcc}
Let $A_m\in\OA(\DK_A), B_n\in\OA(\DK_B); m,n=1,2$ be projections defined in (\ref{A_m})-(\ref{B_n}) with ${\bf a^m}$ and ${\bf b^n}$ given in (\ref{a1})-(\ref{b2}). The correlations $\{(A_m, B_n); m,n=1,2\}$ in the state (\ref{rho}) do \textit{not} have a \textit{commuting} joint common cause $\{C_1, C_2\}$ in $\OA$.
\end{Prop}

Proposition \ref{NCCCCS} answers the question raised at the end of the last Section as to whether there is a common causal justification of the CH inequalities in the general, that is in the noncommuting case. The answer to this question is clearly \textit{no}. The violation of the CH inequality for a given set of correlation does \textit{not} prevent us from finding a common causal explanation for them. All we have to do is to extend our scope of search and to embrace noncommuting common causes in the common causal explanation. So the Bell inequalities in the non-classical case do \textit{not} play the same role as in the classical one. In the classical case there was a direct logical link between the possibility of a common causal explanation and the validity of the Bell inequalities; here the violation of the Bell inequalities excludes only a subset of the possible common causal explanations containing the commuting ones. To put it differently, taking seriously the ontology of AQFT where events are represented by not necessarily commuting projections, one can provide a common causal explanation in a much wider range than simply sticking to commutative common causes.

\section{On the meaning of noncommuting common causes} 

But what are the consequences of applying noncommutative common causes? Let us see the story from the beginning, going back to Reichenbach's original definition of the common cause. The Reichenbachian common cause has the nice property that the presence of a common cause implies a (positive) correlation between the events in question. This fact is a simple consequence of the following identity:
\begin{eqnarray}\label{cons} 
p(A \wedge B)-p(A)\,p(B) = p(C)p(C^{\bot})\big[p(A\vert C)-p(A\vert C^{\bot})\big]\big[p(B\vert C)-p(B\vert C^{\bot})\big] .
\end{eqnarray}
It is straightforward to check that if $C$ is a Reichenbachian common cause fulfilling requirements (\ref{cc1})-(\ref{cc4}) then the right hand side of (\ref{cons}) is positive therefore there is a positive correlation between $A$ and $B$. In this sense the common cause provids a Hempelian explanation for the correlation.\footnote{One is tempted to speculate that this desired property might just have been the reason why Reichenbach took up the statistical relevancy conditions (\ref{cc3})-(\ref{cc4}) in the definition of the common cause.} Going over to the notion of the  common cause system this `explanatory force' of the common cause disappears: from the presence of the common cause (\ref{ccs1}) the correlation  (\ref{classicalcorr}) between $A$ and $B$ does \textit{not} follow. (For an attempt to define the notion of the common cause system such that it preserves this deductive relation between the common cause system and the correlation see (Hofer-Szab\'o and R\'edei 2004, 2006).)

The noncommutative generalization of the common cause system is one step further into the direction of relaxing the relation between the common cause and the correlation. Here not only the deductive relation between the common cause and the correlation gets lost, but also the relation between the conditioned and unconditioned probalitity of the correlating events. Namely, 
\begin{eqnarray}\label{qTB'}
\phi (A) = \phi_c(A):=(\phi\circ E_c)(A)\equiv\sum_k \frac{(\phi\circ E_c)(AC_k)}{\phi(C_k)}\,\phi(C_k)
\end{eqnarray}
holds in general iff $A=E_c(A)$, that is iff $[A, C_k]=0$ for all $k\in K$. That is the state $\phi_c$ differs from $\phi$ for $A\in\OA\setminus {\rm{Im}}\, E_c$ in general, which means that the statistics of $A$ can differ depending on whether we calcutate it directly from the state $\phi$ or as a weighted average of conditional probabilities over the subensembles $C_k$.

But then one might come up with the following concern: Noncommuting common causes are \textit{not actual} but only \textit{contrafactual} entities since if the $C_k$-s \textit{had been realized}, then we \textit{would have ended up} with another probability (the right hand side of (\ref{qTB'})) for the correlating events than the actual ones (the left hand side of (\ref{qTB'})). So these common causes cannot be realized in the same (actual) world in which those event are accomodated which they are supposed to explain.

We do not consider this objection to be serious against the application of noncommuting common causes. An analogy between the notion of the  \textit{common cause} and the notion of the \textit{cause} in QM might help to illuminate why. An observable/event $X$ can be said to be the \textit{cause} of another observable/event $Y$ in QM, if $X$ evolves in time into $Y$. But if $X$ and $Y$ do not commute, then had $X$ been earlier realized, the unitary dynamics would have been distorted, so $X$ would not have evolved into $Y$. Still, we regard $X$ to be the cause of $Y$. Similarly, $C$ is a \textit{common cause} of $A$ and $B$ if conditioned on it the correlation between $A$ and $B$ disappears. If $C$ does not commute with $A$ and $B$, then had $C$ been realized, the statistics would have been distorted, so the probability of $A$, $B$ and $AB$ would be different. Still, we think that $C$ is the common cause.

What is important to see here is that the definition of the common cause does \textit{not} contain the requirement (which our classicaly informed intuition would dictate) that the conditional probabilites, when added up, should give back the unconditional probabilities, that is $\phi=\phi_c$ should fulfil. Or in other words, that the probability of the correlating events should be built up from a finer description of the situation provided by the common cause. To put it in a more formal way: the theorem of total probability is \textit{not} part of the definition of the common cause.\footnote{As it is not part of the definition of the \textit{cause} either: if one measures $X$, one cannot reconstruct the probability of a noncommuting $Y$ from the conditional probabities over the subensembles pertaining to the outcomes of $X$.} The defining property of the common cause is simply the \textit{screening-off}.

So common causes might not be realized without the distortion of the statistics of the original correlating events. But this fact is ubiquitous for noncommuting observables in QM. If we tolerate this fact in general, then why not to tolerate it for common causes? As we have seen, allowing noncommuting common causes helps us to maintain Bell's original intuition concerning local causality.

\section{Conclusions}

In the paper we saw that the Bell inequalities used in AQFT cannot be given a common causal justification similar to the classical Bell inequalities if we allow noncommuting common causes in the explanation. Just the opposite is true: for a set of correlations  violating the CH inequalities a noncommutative common causal explanation can be given and this common cause can be localized in the common past of the correlating events. Thus, abandoning commutativity gives us extra freedom in the search of common causes for correlations. But how big is this freedom? Is it big enough to find a common cause for \textit{any set} of correlations? We saw that for the worst candidate, so to say, for the set \textit{maximally} violating the CH inequality we have found such a common cause. But does it mean that this strategy can be applied across the board? What is the range of correlations for which a joint common causal explanation can be given? Is this range determined only by the size of the set of correlations or by some other properties thereof? Is it true for example that for any finite set of correlations a weak joint common causal explanation can always be given? Or to put it in a more formal way, can one always find a partition of the unit for any finite set of correlations such that the necessary condition (\ref{qCH_mn_noncomm}) for a joint common causal explanation fulfills? All these questions are still open.

\section*{Appendix: In what sense non-classical joint common cause systems are local and non-conspiratorial?}

In Section \ref{non-class} we claimed that Definition \ref{qcccs} of the joint common cause system is the correct non-classical generalization of Definition \ref{lnccccs} of the (classical) \textit{local, non-conspiratorial} joint common cause system. But how can the single non-classical screening-off condition (\ref{qcccs1}) generalize not only the classical screening-off condition (\ref{lncccs1}) but also the locality conditions (\ref{lncccs2a})-(\ref{lncccs2b}) and non-conspiracy (\ref{lncccs3})? This is the question we address in this Appendix.

Let us first introduce a classical probability measure $p_{C_k}$ on a common measure space $(\Omega, \Sigma)$ for every element of a classical common cause system $\{C_k, k\in K\}$, if $p(C_k)\neq 0$:
\begin{eqnarray}\label{notation} 
p_{C_k}(X\vert x) &:=& \frac{p(X \wedge C_k \vert x)}{p(C_k)}.
\end{eqnarray}
With this denotation screening-off (\ref{lncccs1}), locality (\ref{lncccs2a})-(\ref{lncccs2b}), and no-conspiracy (\ref{lncccs3}) will read as
\begin{eqnarray}
p_{C_k}(A_m \wedge B_n \vert a_m \wedge b_n) &=& p_{C_k}(A_m \vert a_m \wedge b_n)\,p_{C_k}(B_n \vert a_m \wedge b_n), 
\label{lncccs1'}\\
p_{C_k}(A_m \vert a_m \wedge b_n) &=& p_{C_k}(A_m \vert a_m \wedge b_{n'}), 
\label{lncccs2a'} \\ 
p_{C_k}(B_n \vert a_m \wedge b_n) &=& p_{C_k}(B_n \vert a_{m'} \wedge b_n), \label{lncccs2b'}\\
p_{C_k}(\Omega\vert a_m \wedge b_n)&=&1, \label{lncccs3'} 
\end{eqnarray}
if one uses no-conspiracy (\ref{lncccs3}) in the first three equations.
The subscript $C_k$ of the probability measure might remind the reader to the standard hidden variable approach where a parameter $\lambda$ is used to index a set of probability measures on a common event algebra. In this approach the derivation of the Bell inequalities then proceeds through the summation/integration over this parameter. In our opinion this indexical treatment of the common cause conceals an important fact, namely that the common cause and the correlating events stand on the same ontological footing: they are all \textit{events}, accomodated in a common event algebra with a single probability measure. Therefore the index in (\ref{lncccs1'})-(\ref{lncccs3'}) is simply an abbreviation of the conditionalization (\ref{notation}), which abbreviation is motivated by trying to find a classically equivalent form, where the non-classicaly meaningless expression $a_m \wedge b_n \wedge C_k$ of non-commuting quantities can have a definite interpretation. (See below.)

Now, how does the non-classical Definition \ref{qcccs} of the joint common cause system relate to the above characterization of a classical \textit{local, non-conspiratorial} joint common cause system? The link is provided by the (in our oppinion) correct interpretation of the non-classical probabilities according to which quantum probabilities are \textit{classical conditional probabilities}. The quantum probability $\phi(X)$ of a projection $X$ is to be interpreted as a \textit{conditional} probability $p(X_{cl} \vert x_{cl})$ of getting the outcome $X_{cl}$ \textit{given} the quantity $x_{cl}$ has been set to be measured. The precise mathematical formulation of this interpretation is given in the so-called `Kolmogorovian Censorship Hyptothesis'. Here we just state the proposition; for the proof see (Bana and Durt 1997), (Szab\'o 2001) and (R\'edei 2010). 
\vspace{0.1in}

\noindent
\textbf{Kolmogorovian Censorship Hypothesis.} Let $(\mathcal{N}, \PR(\mathcal{N}), \phi)$ be a non-classical probability space. Let $\Gamma$ be a countable set of non-commuting selfadjoint operators in $\mathcal{N}$. For every $Q \in \Gamma$, let $\PR(Q)$ be a maximal Abelian sublattice of $\PR(\mathcal{N})$ containing all the spectral projections of $Q$. Finally, let a map $p_0 : \Gamma \rightarrow [0,1]$ be such that
\begin{eqnarray} \label{p_0} 
\sum_{Q \in \Gamma}p_0 (Q) =1, \, \, \, \, \, \,\, \, \,\, \, \,p_0 (Q) > 0.
\end{eqnarray}
Then there exists a classical probability space $(\Omega, \Sigma,p)$ such that for every projection $X^Q$ in any $\PR(Q)$ there exist events $X_{cl}^Q$ and $x_{cl}^Q$ in $\Sigma$ such that
\begin{eqnarray}
X_{cl}^Q \subset x_{cl}^Q \label{cond1} \\
x_{cl}^Q \cap x_{cl}^R = 0, \, \, \, \, \mbox{if} \, \, Q\neq R \label{cond2} \\
p(x_{cl}^Q) = p_0 (Q) \label{cond3} \\
\phi(X^Q) = p(X_{cl}^Q \vert x_{cl}^Q) \label{cond4}
\end{eqnarray}

The intuitive content of the above proposition is the following. A set of incompatible observables represented by noncommuting selfadjoint operators in the set $\Gamma$ are selected for measurement with the probabilities $p_0(Q)$ specified in (\ref{p_0}). This measurement and selection procedure is then represented by classical events $X_{cl}^Q$ and $x_{cl}^Q$, respectively: $X_{cl}^Q$ represents a certain measurement outcome of the measurement $Q$, and $x_{cl}^Q$ is the classical event of setting up the measurement device to measure $Q$. Condition (\ref{cond1}) expresses that no outcome is possible without this setting up of a measuring device. Condition (\ref{cond2}) expresses that incompatible observables $Q$ and $R$ cannot be simultaneously measured: the measurement choices $x_{cl}^Q$ and $x_{cl}^R$ are disjoint events. Condition (\ref{cond3}) states that the classical probability model captures the prescribed probabilities $p_0(Q)$ as the probability of the measurement choices. Finally, condition (\ref{cond4}) is the central relation of the Hypothesis, it states that quantum probabilities can be written as classical conditional probabilities: conditional probabilities of outcomes of measurements on condition that the appropriate measuring device has been set up.
\vspace{0.1in}

\noindent
Applying the above proposition to our case,\footnote{From now on, we will denote both the classical event and the projection representing it by the same symbol. However, the quantum state $\phi$ or the classical probability $p$ will always indicate in which sense we use it.} we obtain that the quantum probabilities $\phi(A_m)$, $\phi(B_n)$ and $\phi(A_m B_n)$ can be interpreted as classical conditional probabilities $p(A_m \vert a_m)$,  $p(B_n \vert b_n)$ and $p(A_m\wedge B_n \vert a_m\wedge b_n)$, respectively, with $A_m, B_n, a_m$ and $b_n$ ($m\in M, n \in N$) accomodated in a classical probability space $(\Omega, \Sigma,p)$. Hence the quantum correlations 
\begin{eqnarray} \label{qqcorr}
\phi(A_m B_n) &\neq& \phi(A_m)\, \phi(B_n)
\end{eqnarray}
between the elements of the set $\{(A_m, B_n); m \in M,n \in N\}$ can be interpreted as \textit{conditional} correlations 
\begin{eqnarray} \label{cccorr} 
p(A_m\wedge B_n\, \vert \, a_m\wedge b_n) &\neq & p(A_m\vert a_m)\, p(B_n \vert b_n)
\end{eqnarray}
between classical measurement outcome events conditioned on measurement choice events in accordance with (\ref{corrcond}).

To see the link between the classical and non-classical version of the common cause let us first introduce a similar notation for the conditionalization on $C_k$ in the non-classical case, if $\phi(C_k) \neq 0$, as was introduced above in (\ref{notation}) for the classical case, that is let
\begin{eqnarray}\label{noncommcond}
\phi_{C_k} (X) &:=& \frac{(\phi\circ E_c)(XC_k)}{\phi(C_k)}= \frac{\phi(C_kXC_k)}{\phi(C_k)}.
\end{eqnarray}
With this notation the definition of the non-classical joint common cause system reads as follows:
\begin{eqnarray}\label{screen'} 
\phi_{C_k} (A_m B_n)&=& \phi_{C_k} (A_m) \, \phi_{C_k} (B_n).
\end{eqnarray}
Using the Kolmogorovian Censorship Hypothesis the classical interpretation of (\ref{screen'}) is the following:
\begin{eqnarray}\label{almostscreen} 
p_{C_k}(A_m\wedge B_n \vert a_m\wedge b_n) &=& p_{C_k}(A_m \vert a_m)\,p_{C_k}(B_n \vert b_n)
\end{eqnarray} 
which is \textit{almost} the screening-off (\ref{lncccs1'}) except that the conditions on the right hand side are \textit{not} $a_m \wedge b_n$. This defect will be cured however by the locality conditions. Observe namely that since $A_m$ and $B_n$ commute, therefore 
\begin{eqnarray}
\phi_{C_k}(A_m) &=& \phi_{C_k}(A_m B_n) + \phi_{C_k}(A_m B^{\perp}_n)\\
\phi_{C_k}(B_n) &=& \phi_{C_k}(A_m B_n) + \phi_{C_k}(A^{\perp}_m B_n)
\end{eqnarray}
which translated into classical conditional probabilities due to the Kolmogorovian Censorship Hypothesis read as:
\begin{eqnarray}
p_{C_k}(A_m\vert a_m) &=& p_{C_k}(A_m \wedge B_n \vert a_m \wedge b_n) + p_{C_k}(A_m \wedge B^{\perp}_n \vert a_m \wedge b_n) = p_{C_k}(A_m \vert a_m \wedge b_n) \label{lncccs2a''} \\
p_{C_k}(B_n\vert b_n) &=& p_{C_k}(A_m \wedge B_n \vert a_m \wedge b_n) + p_{C_k}(A^{\perp}_m \wedge B_n \vert a_m \wedge b_n) = p_{C_k}(B_n \vert a_m \wedge b_n) \label{lncccs2b''}
\end{eqnarray}
Now, observe that (\ref{lncccs2a''})-(\ref{lncccs2b''}) are equivalent to locality (\ref{lncccs2a'})-(\ref{lncccs2b'}), so locality is `automatically' fulfilled for the non-classical common cause due to the commutativity of $A_m$ and $B_n$. (This fact is sometimes referred as the `no-signalling theorem'; for more on that see (Schlieder 1969).) Moreover (\ref{lncccs2a''})-(\ref{lncccs2b''}) also cure the defect of (\ref{almostscreen}), since 
\begin{eqnarray*} 
p_{C_k}(A_m \vert a_m)\,p_{C_k}(B_n \vert b_n)
\end{eqnarray*} 
on the right hand side of (\ref{almostscreen}) can be replaced with 
\begin{eqnarray*} 
p_{C_k}(A_m \vert a_m \wedge b_n)\,p_{C_k}(B_n \vert a_m \wedge b_n)
\end{eqnarray*}
turning (\ref{almostscreen}) into the classical screening-off property (\ref{lncccs1'}). 

Putting all this together, a \textit{non-classica}l, local, non-conspiratorial joint common causal explanation of the correlations (\ref{qqcorr}) is a partition $\left\{ C_k \right\}_{k\in K}\subset\mathcal{P}(\mathcal{N})$ if for any $k\in K$ the following requirements hold:
\begin{eqnarray}
\phi_{C_k} (A_m B_n) &=& \phi_{C_k} (A_m) \, \phi_{C_k} (B_n) \label{qlncccs1}  \\
\phi_{C_k}(A_m) &=& \phi_{C_k}(A_m B_n) + \phi_{C_k}(A_m B^{\perp}_n) \label{qlncccs2a} \\
\phi_{C_k}(B_n) &=& \phi_{C_k}(A_m B_n) + \phi_{C_k}(A^{\perp}_m B_n) \label{qlncccs2b} \\
\phi_{C_k} (\UN) &=& 1.\label{qlncccs3}
\end{eqnarray} 
which using the Kolmogorovian Censorship Hypothesis as a `translation manual' leads us over to the \textit{classical}, local, non-conspiratorial joint common causal explanation (\ref{lncccs1'})-(\ref{lncccs3'}) of the correlations (\ref{cccorr}). But recall that (\ref{qlncccs2a})-(\ref{qlncccs3}) representing locality and no-conspiracy are just \textit{identities}, and hence the screening-off condition (\ref{qlncccs1}) carries the whole content of the common causal explanation---in accordance with our Definition \ref{qcccs}.
\vspace{0.2in}

\noindent
{\bf Acknowledgements.} We wish to thank Jeffrey Bub, John D. Norton, David Malament, Jim Woodward, the Southern California Philosophy of Physics Group and the Foundations of Physics Discussion Group at the University of Maryland for helpful comments on an earlier version of the paper. This work has been supported by the Hungarian Scientific Research Fund, OTKA K-68195 and OTKA K-100715 and by the Fulbright Research Grant while G. H-Sz. was a Visiting Fellow at the Center for Philosophy of Science in the University of Pittsburgh.

\section*{References} 
\begin{list}
{ }{\setlength{\itemindent}{-15pt}
\setlength{\leftmargin}{15pt}}

\item G. Bacciagaluppi, ''Separation theorems and Bell inequalities in algebraic QM,'' \textit{Symposium on the Foundations of Modern Physics 1993: Quantum Measurement, Irreversibility and Physics of Information}, World Scientific, 29-37 (1994).

\item G. Bana and T. Durt, ''Proof of Kolmogorovian Censorship,'' \textit{Foundations of Physics}, \textbf{27}, 1355–1373. (1997). 

\item J. S. Bell, ''Beables for quantum field theory,'' (TH-2053-CERN, presented at the Sixth GIFT Seminar, Jaca, 2–7 June 1975) reprinted in J. S. Bell, \textit{Speakable and Unspeakable in Quantum Mechanics}, (Cambridge: Cambridge University Press, 1987, 52-62.).

\item I. Bengtson and K. \.Zyczkowski, \textit{Geometry of Quantum States: An Introduction to Quantum Entanglement}, Cambridge University Press, Cambridge, 2006.

\item J. Butterfield, ''Vacuum correlations and outcome independence in algebraic quantum field theory'' in D. Greenberger and A. Zeilinger (eds.), \textit{Fundamental Problems in Quantum Theory, Annals of the New York Academy of Sciences, Proceedings of a conference in honour of John Wheeler}, 768-785 1995.

\item D. Buchholz, C. D'Antoni and K. Fredenhagen, ''The universal structure of local algebras,'' \textit{Commun. Math. Phys.}, \textbf{111/1}, 123-135 (1987). 

\item J. F. Clauser, M.A. Horne, A. Shimony and R. A. Holt, ''Proposed experiment to test local hidden-variable theories,'' \textit{Phys. Rev. Lett.}, \textbf{23}, 880-884 (1969).

\item J. F. Clauser and M. A. Horne, ''Experimental consequences of objective local theories,'' \textit{Phys. Rev. D}, \textbf{10}, 526-535 (1974).

\item R. Clifton and L. Ruetsche, ''Changing the subject: Redei on causal dependence and screening off in relativistic quantum field theory,'' \textit{Philosophy of Science}, \textbf{66}, S156-S169 (1999).

\item K. Fredenhagen, ''On the modular structure of local algebras of observables'' \textit{Commun. Math. Phys.}, \textbf{97}, 79-89 (1985). 

\item N. Gisin, \textit{Quantum Reality, Relativistic Causality, and Closing the Epistemic Circle}, (The Western Ontario Series in Philosophy of Science, \textbf{73}, III / 1, 125-138, 2009).

\item C. Glymour, ''Markov properties and quantum experiments,'' in W. Demopoulos and I. Pitowsky (eds.) \textit{Physical Theory and its Interpretation}, (Springer, 117-126, 2006).

\item R. Haag, {\it Local Quantum Physics}, (Springer Verlag, Berlin, 1992). 

\item H. Halvorson and R. Clifton, ''Generic Bell correlation between arbitrary local algebras in quantum field theory,''
\textit{J. Math. Phys.}, \textbf{41}, 1711-1717 (2000).

\item H. Halvorson, ''Algebraic quantum field theory,'' in J. Butterfield, J. Earman (eds.), \textit{Philosophy of Physics, Vol. I}, Elsevier, Amsterdam, 731-922 (2007).

\item J. Henson (2005). "Comparing causality principles," {\it Studies in the History and Philosophy of Modern Physics}, {\bf 36}, 519–543.

\item G. Hofer-Szab\'o, G., M. R\'edei (2004). "Reichenbachian Common Cause Systems," {\it International Journal of Theoretical Physics}, {\bf 34}, 1819-1826.

\item G. Hofer-Szab\'o, G., M. R\'edei (2006). "Reichenbachian Common Cause Systems of Arbitrary Finite Size Exist," {\it Foundations of Physics Letters}, {\bf 35}, 745-746.

\item G. Hofer-Szab\'o and P. Vecserny\'es ''Reichenbach's Common Cause Principle in AQFT with locally finite degrees of freedom,'' \textit{Found. Phys.}, \textbf{42}, 241-255 (2012a).

\item G. Hofer-Szab\'o, P. Vecserny\'es, ''Noncommutative Common Cause Principles in AQFT,'' \textit{Found. Phys.}, (submitted) (2012b).

\item G. Hofer-Szab\'o, P. Vecserny\'es, ''Noncommuting local common causes for correlations violating the Clauser--Horne inequality,'' \textit{Journal of Physics A}, (submitted) (2012c).

\item J. Jarrett, ''On the Physical Significance of the Locality Conditions in Bell Arguments,'' \textit{Nous}, \textbf{18}, 569-589, (1984).

\item L. Landau, ''On the violation of Bell's inequality in quantum theory,'' \textit{Phys. Lett. A} \textbf{120}, 54-56 (1987).

\item G. L\"uders, ''\"Uber die Zustands\"anderung durch den Messprozess,'' \textit{Annalen der Physik} \textbf{8}, 322-328 (1951)..

\item V.F. M\"uller and P. Vecserny\'es, ''The phase structure of $G$-spin models'', \textit{to be published}

\item F. Nill and K. Szlach\'anyi, ''Quantum chains of Hopf algebras with quantum double cosymmetry'' \textit{Commun. Math. Phys.}, \textbf{187} 159-200 (1997).

\item M. R\'edei, ''Reichenbach's Common Cause Principle and quantum field theory,'' \textit{Found. Phys.}, \textbf{27}, 1309--1321 (1997).

\item M. R\'edei, {\it Quantum Logic in Algebraic Approach}, (Kluwer Academic Publishers, Dordrecht, 1998).

\item M. R\'edei, ''Kolmogorovian Censorship Hypothesis for general quantum probability theories,'' \textit{Manu\-scrito - Revista Internacional de Filosofia}, \textbf{33}, 365–380 (2010).

\item M. R\'edei and J. S. Summers, ''Local primitive causality and the Common Cause Principle in quantum field theory,'' \textit{Found. Phys.}, \textbf{32}, 335-355 (2002).

\item M. R\'edei and J. S. Summers, ''Remarks on Causality in relativistic quantum field theory,'' \textit{Int. J. Theor. Phys.}, \textbf{46}, 2053–2062 (2007).

\item H. Reichenbach, {\it The Direction of Time}, (University of California Press, Los Angeles, 1956).

\item A. Shimony, ''Events and processes in the quantum world,'' in Penrose, R. and Isham, C. (eds.), \textit{Quantum concepts in space and time}, 182–203. (University Press, Oxford, 1986.)

\item S. Schlieder, ''Einige Bemerkungen \"uber Projektionsoperatoren (Konsequenzen eines Theorems von Borchers,'' \textit{Communications in Mathematical Physics}, \textbf{13},  216-225 (1969).

\item S. J. Summers, ''On the independence of local algebras in quantum field theory'' \textit{Reviews in Mathematical Physics}, \textbf{2}, 201-247 (1990).

\item S. J. Summers and R. Werner, ''Bell's inequalities and quantum field theory, I: General setting,'' \textit{Journal of Mathematical Physics}, \textbf{28}, 2440-2447 (1987a).

\item S. J. Summers and R. Werner, ''Bell's inequalities and quantum field theory, II: Bell's inequalities are maximally violated in the vacuum,'' \textit{Journal of Mathematical Physics}, \textbf{28},  2448-2456 (1987b).

\item S. J. Summers and R. Werner, ''Maximal violation of Bell's inequalities for algebras of observables in tangent spacetime regions,'' \textit{Ann. Inst. Henri Poincar\'e -- Phys. Th\'eor.}, \textbf{49}, 215-243 (1988).

\item L. E. Szab\'o, ''Quantum structures do not exist in reality,'' \textit{Int. J. of Theor. Phys}, \textbf{37}, 449-456. (1998)

\item L. E. Szab\'o, ''Critical reflections on quantum probability theory,'' in R\'edei and St\"olzner (eds.) \textit{John von Neumann and the Foundations of Quantum Physics}. (Institute Vienna Circle Yearbook. Dordrecht: Kluwer Academic Publishers, 201–219, 2001)

\item K. Szlach\'anyi and P. Vecserny\'es, ''Quantum symmetry and braid group statistics in $G$-spin models'' \textit{Commun. Math. Phys.}, \textbf{156}, 127-168 (1993). 

\item H. Umegaki, ''Conditional expectation in an operator algebra, I,'' \textit{Tohoku Math. J. (2)}, \textbf{6/2-3}, 177-181 (1954).

\item B. C. Van Fraassen, ''Rational belief and common cause principle,'' in: R. McLaughlin (ed.), {\it What? Where? When? Why?}, Reidel, 193-209 (1982).

\item R. M. Wald, ''Correlations beyond the horizon,'' {\it General Relativity and Gravitation},  \textbf{24}, 1111-1116 (1992).

\end{list}
\end{document}